\documentclass[epj]{svjour}
\usepackage{epsf}
\begin{document}
\date{}
\title{Andreev-Lifshitz supersolid revisited 
for a few electrons on a square lattice II}
\titlerunning{Andreev-Lifshitz supersolid II}

\author{Zolt\'an \'Ad\'am N\'emeth$^{(a,b)}$ and Jean-Louis 
Pichard$^{(a,c)}$}

\authorrunning{Z. \'A. N\'emeth and J.-L. Pichard}
        
\institute{
(a) CEA/DSM, Service de Physique de l'Etat Condens\'e, 
Centre d'Etudes de Saclay, 91191 Gif-sur-Yvette Cedex, France \\  
(b) E\"otv\"os University, Departement of Physics of Complex Systems, 
1117 Budapest, P\'azm\'any P\'eter s\'et\'any 1/A, Hungary \\
(c) Laboratoire de Physique Th\'eorique et Mod\'elisation, 
Universtit\'e de Cergy-Pontoise, 95031, Cergy-Pontoise Cedex, France 
}

\abstract
{
In this second paper, using $N=3$ polarized electrons (spinless fermions) 
interacting via a $U/r$ Coulomb repulsion on a two dimensional $L \times L$ 
square lattice with periodic boundary conditions and nearest neighbor 
hopping $t$, we show that a single unpaired fermion can co-exist with a 
correlated two particle Wigner molecule for intermediate values of the 
Coulomb energy to kinetic energy ratio $r_s=UL/(2t\sqrt{\pi N})$. This 
supports in an ultimate mesoscopic limit a possibility proposed by Andreev 
and Lifshitz for the thermodynamic limit: a quantum crystal may have 
delocalized defects without melting, the number of sites of the crystalline 
array being smaller than the total number of particles. When $L=6$, the 
ground state exhibits four regimes as $r_s$ increases: a Hartree-Fock regime, 
a first supersolid regime where a correlated pair co-exists with a third 
fully delocalized particle, a second supersolid regime where the third 
particle is partly delocalized, and eventually a correlated lattice regime. 
} 

\PACS{
{71.10.-w}   Theories and models of many-electron systems \and 
{73.21.La}    Quantum dots  \and 
{73.20.Qt}   Electron solids  
} 

\maketitle

\section{Introduction}

 In 1969, it was conjectured by Andreev and Lifshitz \cite{andreev-lifshitz} 
that at zero temperature, delocalized defects may exist in a quantum solid, 
as a result of which the number of sites of an ideal crystal lattice may not 
coincide with the total number of particles. Originally, this conjecture 
was proposed for three dimensional quantum solids made of atoms 
($He^3$, $He^4$, $\ldots$) which do not interact via Coulomb 
repulsion. We re-visit such a possibility for electron solids 
with long range Coulomb repulsion in two dimensions. The motivation 
to re-visit nowadays this issue comes from questions raised by the 
physics of electrons in Si MOSFETs and similar $2d$ field effect devices. 
An unexpected low temperature metallic behavior \cite{abrahams} has been 
observed at intermediate values of the Coulomb energy to kinetic energy ratio 
$r_s$, which remains unexplained. Another actual motivation is given by 
the promising perspectives opened by trapped cold ion systems, where 
one can study how a Wigner molecule becomes a quantum fluid when the 
ions are squeezed \cite{dubin}.
 
 In a first paper \cite{ksp}, the supersolid phase conjectured 
\cite{andreev-lifshitz} by Andreev and Lifshitz was introduced, together 
with a related variational approach using a fixed number of fermions BCS 
wave function \cite{bouchaud} of Bouchaud et al. The question 
is to know if a system of unpaired electrons with a reduced Fermi energy 
can co-exist with an ordered array of charges, the number of sites of the 
crystalline array being smaller than the total number of electrons. In 
Ref. \cite{ksp}, this question was investigated using $N=4$ spinless 
fermions interacting via a $U/r$ Coulomb repulsion in a two dimensional 
$6 \times 6$ square lattice with periodic boundary conditions (BCs) 
and nearest neighbor hopping $t$. It was observed that for intermediate 
ratios $r_s=UL/(2t\sqrt{\pi N})$ (typically for $10 < r_s < 28$), 
the ground state (GS) is in a mixed state, where unpaired delocalized 
fermions co-exist with a strongly paired, nearly solid assembly. From the 
study of the different inter-particle spacings as $r_s$ increases, it was 
concluded that before having full Wigner crystallization, a floppy three 
particle Wigner molecule is formed, while the fourth particle remains 
delocalized. We consider in this second paper the ultimate limit $N=3$, 
where it is still possible to exactly study if a correlated two particle 
molecule can co-exist with a third delocalized particle. As in the first 
paper, the study is restricted to fully polarized electrons (spinless 
fermions) having anti-symmetric orbital wave functions. A study involving 
the spin degrees of freedom can be found in Ref. \cite{selva}. 

 The paper is organized as follows. Once the lattice model is defined 
in Sec. \ref{section1}, the three regimes characterizing the formation 
of a two particle Wigner molecule (2PWM) on an empty periodic lattice are 
summarized in Sec. \ref{section2} and in App. \ref{Appendix A}: a weak 
coupling Fermi regime, a correlated Wigner regime with harmonic oscillatory 
motions of the particles around equilibrium, and a correlated lattice regime 
where these oscillatory motions become damped by the lattice. 
The weak coupling Fermi limit and the strong coupling correlated lattice 
limit of the three particle system are described in Sec. \ref{section3} 
and Sec. \ref{section4}. An additional discussion of the correlated lattice 
limit when $L \rightarrow \infty$ is given in App. \ref{Appendix B} both 
for the zero density limit (keeping $N=3$) and for the constant density 
limit (taking $N=L^2/9-1$). The threshold  $r_s^F$ above which a weak 
coupling expansion in powers of $U/t$ and the threshold $r_s^*$ under 
which a strong coupling lattice expansion in powers of $t/U$ cease 
to be valid are defined in Sec. \ref{section5}. Taking $L=6$, one gets 
the range of intermediate values $r_s^F \approx 6 < r_s < r_s^* \approx 180$ 
where the GS structure is non-trivial. A simple ansatz for the GS wave 
function, first introduced in Ref. \cite{nemeth}, is studied in 
Sec. \ref{section6}, corresponding to a 2PWM co-existing with a third 
particle which remains partly delocalized in the direction parallel to 
the 2PWM. The ansatz combines two possible directions for the 2PWM and 
a delocalized center of mass. This defines the concept of a partially 
melted Wigner molecule (PMWM) near the lattice limit, and describes 
the three particle GS at $r_s \approx 40$, when the 2PWM oscillations 
are taken into account using a lattice $t/U$ expansion. In Sec. 
\ref{section7} we show that when $r_s$ is further decreased, the GS is 
made of a floppy 2PWM with large oscillatory motions which are not damped 
by the lattice, co-existing with a third fully delocalized particle. 
The unpaired particle simply provides a uniform background density for 
the 2PWM. The total momentum ${\bf K}_2$ of the 2PWM and the momentum 
${\bf k}_{3}$ of the third particle satisfy the conservation of the total 
momentum ${\bf K}= {\bf K}_2 + {\bf k}_{3}$. This yields different possible 
combinations which contain more than more than 90 \% of the exact GS when 
$r_s \approx 10$. In Sec. \ref{section9}, we summarize the four regimes 
found for the three particle GS when $L=6$, pointing out the role of the 
lattice and raising the main question: Are those observed supersolid GSs 
the ultimate mesoscopic trace of a thermodynamic supersolid phase proposed 
by Andreev and Lifshitz, consisting of a $N'$ electron solid co-existing with 
a $N-N'$ electron fluid, out of a total number $N$ of electrons? The GS 
nodal structure and the occupation numbers $P_{\bf k}$ on the reciprocal 
lattice are studied in App. \ref{Appendix C} and in App. \ref{Appendix D}.

\section{Lattice model}
\label{section1}

 The Hamiltonian of the $L \times L$ square lattice model with periodic 
BCs is the same as in Ref. \cite{ksp}. Denoting $c_{\vec{j}}^{\dagger}$, 
$c_{\vec{j}}$ the creation, annihilation operators of a spinless fermion 
at the site $\vec{j}$, it reads:   
\begin{equation}
H=-t \sum_{\left<{\bf j},{\bf j'}\right>}c_{\bf j}^\dagger c_{\bf j'} + 
{U\over2} \sum_{{\bf j}, {\bf j'} \atop {\bf j}\neq {\bf j'}}{n_{\bf j}  
n_{\bf j'}\over d_{\bf jj'}}.  
\label{Hamilton1}
\end{equation}
The effective mass being $m^*$, $t=\hbar^2/(2m^* a^2)$ is the hopping 
term between nearest neighbors, and $U=e^2/(\epsilon a)$ is the Coulomb 
interaction between two fermions separated by a lattice spacing $a$ in a 
medium of dielectric constant $\epsilon$. The Coulomb energy to Fermi 
energy ratio $r_s$ becomes in this lattice model 
\begin{equation}
r_s={U\over 2t\sqrt{\pi \nu}}
\end{equation}
where $\nu=N/L^2$.  For $L=6$, this gives $r_s = 0.98 U/t$ if $N=3$ 
and $r_s = 1.20U/t$ if $N=2$.

Without disorder, the Hamiltonian (\ref{Hamilton1}) is more conveniently 
written using the operators $d_{\vec{k}}^{\dagger}$ ($d_{\vec{k}}$) 
creating (annihilating) a spinless fermion in a single particle plane 
wave state of momentum $\vec{k}$. The Hamiltonian (\ref{Hamilton1}) becomes: 
\begin{eqnarray}
H&=&-2t \sum_k \left(\cos k_x +\cos k_y \right)d_{\bf k}^\dagger
d_{\bf k} \nonumber \\ 
&&+U \sum_{{\bf k},{\bf k'},{\bf q}} V({\bf q}) d_{{\bf k}+{\bf q}}^\dagger 
d_{{\bf k'}-{\bf q}}^\dagger d_{\bf k'} d_{\bf k}
\end{eqnarray}
where
\begin{equation}
V({\bf q})={1\over 2 L^2} \sum_{\bf j} {\cos{\bf qj}\over d_{\bf j0}}. 
\label{int}
\end{equation}
The distance $d_{\bf j0}$ is defined as the shortest distance between 
the sites $\bf j$ and $\bf 0$ of the square lattice with periodic BCs:   
\begin{equation}
d_{\bf j0}=\sqrt{\min(j_x,L-j_x)^2+\min(j_y,L-j_y)^2}.
\label{distance}
\end{equation}
The states of different total momenta $\vec{K}$ are decoupled. 
Moreover, since the Coulomb repulsion is a two-body interaction, 
only states of same $\vec{K}$ having $N-2$ $\vec{k}$ in common 
out of $N$ are directly coupled. When $N\geq 3$, this means that 
the Hamiltonian matrix of a subspace of given $\vec{K}$ is sparse.

\section{Formation of a two-particle Wigner molecule on an empty 
square lattice}
\label{section2}

Before studying the three particle problem, we summarize 
the three regimes characterizing the two particle problem
on an empty lattice. Firstly, it allows us to introduce the value 
$r_s^*$ above which the lattice effects become important. Secondly, 
it will be useful for analyzing in Sec. \ref{section6} the three 
particle GS in terms of a two particle Wigner molecule created  
on the uniform background provided by a third delocalized particle.  

Following Ref. \cite{moises}, we consider the relative fluctuations 
\begin{equation}
u_r =\frac{\sqrt{\langle r^2\rangle-\langle r \rangle^2}}{\langle r\rangle} 
=\frac{\Delta r}{\langle r\rangle}
\label{u_r}
\end{equation}
of the distance $r$ between the two particles. This gives three 
regimes: 

\begin{itemize}

\item For $r_s < r_s^F = \pi^{3/2}$, the fluctuation $u_r$  keeps 
essentially its non-interacting value, up to some negligible 
perturbative corrections.

\item For $ r_s^F <r_s < r_s^* \propto L^3$, $u_r$ decays as $r_s^{-\alpha}$, 
the two particles beginning to form a correlated Wigner molecule, with an 
oscillatory motion of the particles around the equilibrium position of 
the molecule. The interaction $U n_{\bf j} n_{\bf j'}/ 2 d_{\bf jj'}$ 
having a cusp at the equilibrium position if one defines $d_{\bf jj'}$ 
as previously, the oscillations are not harmonic and one gets 
$\alpha \approx 0.31$ \cite{moises}. Smearing this cusp, one recovers 
harmonic oscillations, and  $\alpha =1/4$, as shown in App. 
\ref{Appendix A}. This is the harmonic regime first considered by 
Wigner \cite{wigner} (see also Ref. \cite{carr}) for the continuum 
electron gas. 

\item If one subtracts a finite size correction $\approx 0.9/(L+1)^2$ 
($L$ even) from $u_r$, one obtains a universal scaling law 
which depends only on $r_s \approx 0.2 UL/t$ ($N=2$) with a 
crossover from independent particle motion towards correlated 
motion at $r_s^F$. 

\item When $r_s$ reaches a higher threshold $r_s^*$, the oscillations of 
the inter-particle spacing become of the order of the lattice spacing $a$ and 
a lattice expansion in powers of $t/U$ becomes valid. The continuum-lattice 
crossover occurs when $\Delta r \approx 2/3$ (in units of $a$). A $t/U$ 
expansion giving:    
\begin{equation}
\Delta r =\frac{t}{U}\, L\, \sqrt{\Bigl[(L-1)^2+1\Bigr]
\left(3+\cos\frac{2\pi}{L}\right)}
\label{flu}
\end{equation}
one gets a lattice threshold $r_s^* \propto L^3$. For $N=2$ and $L=6$, 
$r_s^* \approx 100$. Above $r_s^*$, $u_r \propto tL/U$ is no longer a 
universal function of the ratio $r_s \propto UL/t$. 
 
\end{itemize}

\begin{figure}
{\centerline{\leavevmode \epsfxsize=9cm \epsffile{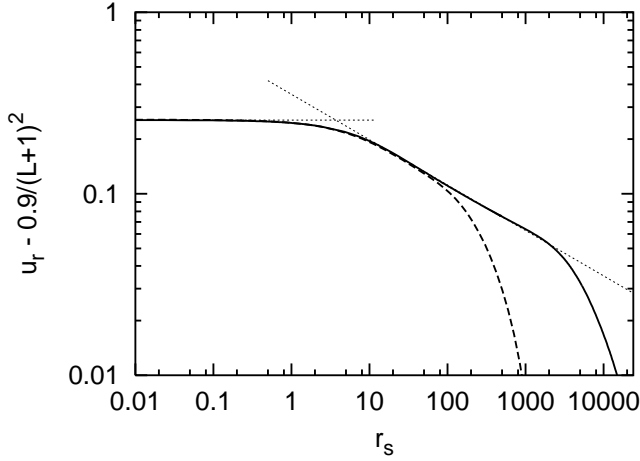}}}
\caption{Relative fluctuation $u_r$ of the inter-particle distance $r$ 
for two particles on an empty $L \times L$ square lattice (minus a finite 
size correction $0.9/(L+1)^2$) as a function of $r_s$ for 
$L=6$ (dashed line) and $L=12$ (continuous line). The smeared interaction 
(Eq. \ref{smeared}) has been taken. The dotted line 
corresponds to the $u_r \propto r_s^{-1/4}$ behavior derived in App.  
\ref{Appendix A}. Notice the breakdown of the universal scaling when 
$r_s > r^*(L)$.}
\label{FIG1}
\end{figure}

 The three behaviors of $u_r$ are given in Fig. 4 of Ref. \cite{moises} 
for various values of $L$ if one takes  the Coulomb interaction which we 
assume in this work. The relative fluctuations $u_r$ yielded by a smeared 
Coulomb interaction (Eq. \ref{smeared} of App. \ref{Appendix A}) are shown in 
Fig. \ref{FIG1} for $L=6$ and $12$. One can see that when $r_s < r_s^*$, 
$u_r-0.9/(L+1)^2$ is a universal function of $r_s$ with a Fermi-Wigner 
crossover at $r_s^F \approx \pi^{3/2}$, and that this universal regime ceases 
to be valid due to lattice effects when $r_s$ exceeds $r_s^*$. 
 
\section{The Fermi limit}
\label{section3}

\begin{figure}[t]
{
\centerline{\leavevmode \epsfxsize=9cm \epsffile{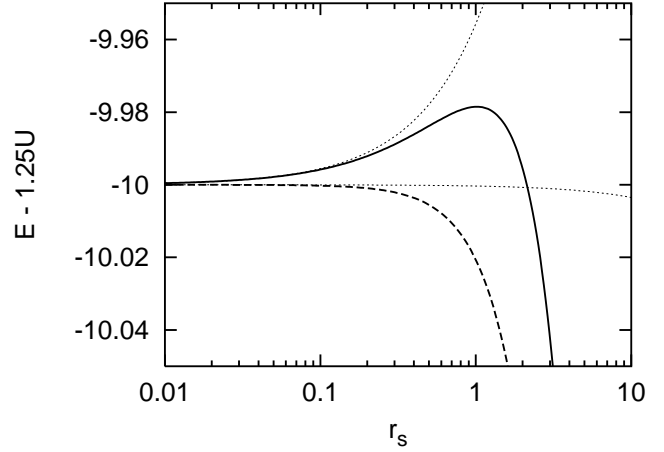}}}	
\caption{The GS energies $E$ for ${\bf K}=(0,0)$ (thick line) 
and ${\bf K}=2\pi/6(1,1)$ (thick dashed line) as a function of 
$r_s$. The thin dotted lines give the behaviors 
$E^{(I)}_{{\bf K}=(0,0)}= -10t+1.2933 U$ and 
$E^{(I)}_{{\bf K}=2\pi/6(1,1)}= -10t+1.2497 U$ obtained at 
the leading order of a $U/t$ expansions (Eqs. \ref{U-expansion1} 
and \ref{U-expansion2}).
}
\label{fig31}
\end{figure}

We begin to study three particles on $6 \times 6$ periodic lattice when 
$U \rightarrow 0$. In this limit, the eigenstates 
$\left|\Psi(r_s=0)\right>$ are plane-wave states: 
\begin{equation}
\left|\Psi(r_s=0)\right>=c^\dagger_{\vec{k}_1} c^\dagger_{\vec{k}_2}
c^\dagger_{\vec{k}_3}\left|0\right>,
\end{equation}
$\left|0\right>$ being the vacuum state. The GS energy $-10t$ has a 
sixfold degeneracy. A basis of this degenerate eigenspace can be built 
using two states of total momentum ${\bf K}=(0,0)$, given by 
\begin{equation} 
c^\dagger_{(0,0)} c^\dagger_{{2\pi\over6}(1,0)} 
c^\dagger_{{2\pi\over 6}(-1,0)}\left|0\right> 
\end{equation} 
and its $x\leftrightarrow y$-symmetric counterpart, and four states 
\begin{equation} 
c^\dagger_{(0,0)} c^\dagger_{{2\pi\over 6}(\pm 1,0)} 
c^\dagger_{{2\pi\over 6}(0,\pm 1)}\left|0\right>
\end{equation} 
of total momenta ${\bf K}=2\pi/6 (\pm 1,\pm 1)$. 
 
When $U/t$ is small, one can use perturbation theory to determine which 
of those six states have the lowest energy when one switches on $U$. 
At first order, the corrections $\Delta E^{(I)}_{\bf K}$ to the GS 
energy $-10t$ are given by the diagonal elements of the interaction 
matrix (the two ${\bf K}=(0,0)$ states being decoupled 
due to the additional $x\leftrightarrow y$ symmetry). 
One gets for ${\bf K}=(0,0)$
\begin{equation}
\frac{\Delta E^{(I)}_{\bf K}}{U} = 6 V(0,0)-4 
V\left({2\pi\over 6},0\right)-2 V\left({4\pi\over 6},0\right)
\label{U-expansion1}
\end{equation}
and for ${\bf K}=({2\pi/ 6},{2\pi/ 6})$: 
\begin{equation}
\frac{\Delta E^{(I)}_{\bf K}}{U} =6 V(0,0)
-4 V\left({2\pi\over 6},0\right)-2 V\left({2\pi\over6},{2\pi\over 6}\right)
\label{U-expansion2}
\end{equation}
where the $V(q_x,q_y)$ are given by Eq. \ref{int}. 

 The sixfold degeneracy is partly removed by $U$, the four 
states of total momenta ${\bf K}=2\pi/6(\pm 1,\pm 1)$ having a smaller energy 
than the two states of $\vec{K}=(0,0)$.  One can compare in Fig. \ref{fig31}
the exact behaviors and the first order perturbative expansions 
(Eqs. \ref{U-expansion1} and \ref{U-expansion2}) for 
${\bf K}=2\pi/6(\pm 1,\pm 1)$ and 
$\vec{K}=(0,0)$ total momenta.

\section{The correlated lattice limit}
\label{section4}

\begin{figure}[b]
{\centerline{\leavevmode \epsfxsize=9cm \epsffile{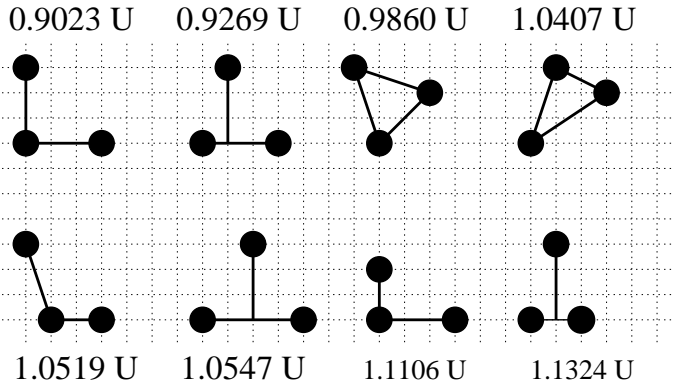}}}	
\caption{Some of the low energy configurations with their
electrostatic energy on $L=6$ lattice.}
\label{fig401}
\end{figure}

\begin{figure}[t]
{\centerline{\leavevmode \epsfxsize=5cm \epsffile{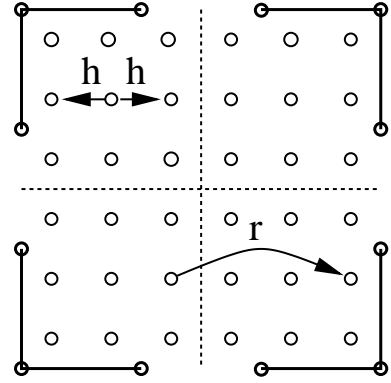}}}	
\caption{The $L^2=36$ triangles of Fig. \ref{fig401} can be put on 
an effective $L \times L$ periodic lattice. The four sectors correspond to 
the four possible orientations of the triangles (shown in the corner), 
each oriented triangle having $L/2 \times L/2$ possible locations on 
the original  $L \times L$ periodic lattice. The first neighbor hops $h$ 
and third neighbor hops $r$ correspond to a $t/U$ expansion at the order 
$N=L/2=3$. 
}
\label{fig40}
\end{figure}

We now consider the limit $t \rightarrow 0$. Without hopping $t$, the 
three particles stay localized  on three different lattice sites, 
forming configurations which can be ordered by increasing Coulomb energy. 
For the low energy configurations, the inter-particle spacings are as large 
as it is possible on a periodic square lattice. The first 
configurations of minimum Coulomb energy are given in Fig. \ref{fig401}. 
Without disorder, the $L^2$ sites are equivalent, and $L^2=36$ identical 
configurations can be put on the $L \times L$ lattice unless an extra 
symmetry or the periodic BCs reduces this number. This yields 
large degeneracies when $t=0$. For instance the GS degeneracy is equal 
to $36$, the states being triangles of Coulomb energy $0.9023 U$, having 
different locations or orientations (see Fig. \ref{fig40}). 

This large degeneracy can be partly broken by a hopping term $t\neq 0$. 
This can be studied using a perturbation theory starting from the $t=0$ 
triangles and taking $t$ as a perturbation. The first correction 
to Coulomb energy of the $L^2$ triangles is given  at the second order. 
One gets a uniform shift $\propto t^2/U$ which does not remove the 
$L^2$ degeneracy. For the $36$ triangles ($\beta=1\dots 36$) of 
energy $E_0^{(0)}=0.9023 U$, one gets the same second order correction:
\begin{equation}
\Delta E_{0\beta}^{(II)}=\sum_{\alpha \atop \alpha\neq
0}{\left<\Psi_{0\beta} \left|H_1\right| \Psi_\alpha \right>
\left<\Psi_\alpha \left|H_1\right| \Psi_{0\beta}\right>\over
(E_0^{(0)}-E_\alpha)}. 
\end{equation}

At the third order, two processes become possible for $N=3$ and $L=6$: 
either the $N=3$ particles hop together by one lattice spacing in the same 
direction, such that the center of mass of the corresponding triangle is 
translated by the same hop (hopping term $h \propto t^N/U^{N-1}$) or one 
particle hops over a scale $L/2=3$ (hopping term $r \propto 
t^{L/2}/U^{L/2-1}$. This $L/2=3$ hop  couples two triangles having in 
common two sites, i.e changes  the orientation of the triangle. Those 
two processes are possible at the same order when $N=L/2$, which is our 
case. Inside the subspace spanned by the $36$ triangles, the translational 
invariance is recovered at the order $N=L/2=3$. Each eigenstate can be now 
labeled by its quantized total momentum ${\bf K}$. This partly removes 
the $36$ degeneracy of the triangles. The matrix elements of 
the $36 \times 36$ secular matrix are given at the order $N=L/2=3$ by: 
\begin{equation}
M_{\beta,\beta'}^{(III)}=\sum_{\alpha,\alpha' \atop \alpha,\alpha'\neq
0}{\left<\Psi_{0\beta} \left|H_1\right| \Psi_\alpha \right>
\left<\Psi_\alpha \left|H_1\right| \Psi_{\alpha'}\right>
\left<\Psi_\alpha \left|H_1\right| \Psi_{0\beta'}\right>\over
(E_0^{(0)}-E_\alpha)(E_0^{(0)}-E_{\alpha'})},
\label{secu3}
\end{equation}
where $\beta$, $\beta'$ labels two different triangles. The diagonalization 
of this matrix is easy, since we can order the $36$ triangles of a $6 \times 
6$ lattice as indicated in Fig. \ref{fig40}. A site $\vec{j}$ 
of this effective lattice corresponds to a triangle, and 
$D^{\dagger}_{\bf j}$ ($D_{\bf j}$) are the corresponding creation 
(annihilation) operators
\begin{equation}
D^\dagger_{\bf j} = c^\dagger_{\bf j} c^\dagger_{(j_x,j_y+3)} 
c^\dagger_{(j_x+3,j_y)}.
\label{cre3}
\end{equation}
One gets the effective Hamiltonian $H_{eff}^{(III)}$: 
\begin{equation}
H_{eff}^{(III)}=\sum_{\bf j} E_0^{(II)} D_{\bf j}^\dagger D_{\bf j}+h
\sum_{\left<{\bf j},{\bf j'}\right>} D_{\bf j}^\dagger D_{\bf j'}+{r \over2}
\sum_{\left<{\bf j},{\bf j'}\right>_3}D_{\bf j}^\dagger D_{\bf j'},
\label{effham3}
\end{equation}
which is identical to a one particle Hamiltonian describing the motion of 
a single particle on a $6 \times 6$ square lattice with periodic BCs, 
with first neighbor hopping matrix element $h$ and third neighbor hopping 
matrix element $r$. $h$ and $r$ are given by the corresponding matrix 
elements of $M_{\beta,\beta'}^{(III)}$.

 For $N=3$ and $L=6$, the eigenenergies of $H_{eff}^{(III)}$ are:
\begin{eqnarray}
\Delta E^{(III)}_{\bf K} &=& E^{(III)}_{\bf K}-{E}_0^{(II)} \\
                  &=& 2h\left(\cos K_x+\cos K_y\right) + r 
\left(\cos 3 K_x + \cos 3 K_y  \right), \nonumber
\end{eqnarray}
the results being summarized in Table \ref{pertres1} at the fourth 
order of a $t/U$ expansion with: 
\begin{equation}
{E}_0^{(IV)} =0.9023U-208.9{t^2\over U}-682883{t^4\over U^3} 
\end{equation} 
and $h=1000 t^3/U^{2}$, $r=3320 t^3/U^{2}$ and $s=61926 t^4/U^{3}$. 

\begin{table}[hbt]
\begin{center}
\begin{tabular}{||c|c||} \hline
${\bf K}\cdot 6/2\pi$&$E_{pert}$ \\ \hline \hline
$(0,0)$&${E}_0^{(IV)}-4h+2r+4s$ \\ \hline
$(1,0)$&${E}_0^{(IV)}-3h+s$ \\ \hline
$(1,1)$&${E}_0^{(IV)}-2h-2r-2s$ \\ \hline
$(2,0)$&${E}_0^{(IV)}-h+2r+s$\\ \hline
$(2,1)$&${E}_0^{(IV)}-2s$ \\ \hline
$(2,2)$&${E}_0^{(IV)}+2h+2r-2s$ \\ \hline
$(3,0)$&${E}_0^{(IV)}+4s$ \\ \hline
$(3,1)$&${E}_0^{(IV)}+h-2r+s$ \\ \hline
$(3,2)$&${E}_0^{(IV)}+3h+s$ \\ \hline
$(3,3)$&${E}_0^{(IV)}+4h-2r+4s$ \\ \hline
\end{tabular}
\end{center}
\caption{The $L^2$ lowest energies given by a $t/U$ expansion 
up to order four. The other ${\bf K}$-states follow from the x-y, 
and the ${\bf K} \rightarrow {\bf -K}$ symmetries.
\label{pertres1}}
\end{table}

\begin{figure}[b]
{\centerline{\leavevmode \epsfxsize=9cm \epsffile{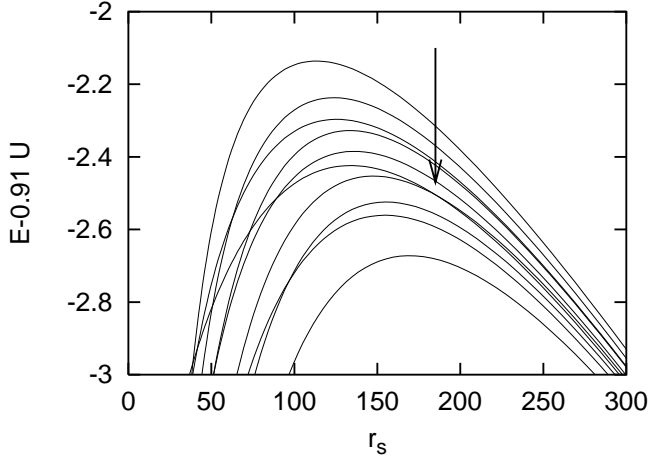}}}	
\caption
{The 36 lowest energies as a function of $r_s$ for all possible total 
momenta $\bf{K}$. Above the last level crossing indicated 
by the arrow the momenta of the 36 first states 
are given by the $t/U$-expansion (Table \ref{pertres1}). 
}
\label{fig41}
\end{figure}

 One can see in Fig. \ref{fig41} that this $t/U$ expansion gives 
the exact momenta $\bf{K}$ of the $36$ first states 
above a last level crossing at $r_s = 196$. For $L=6$, one has 
no GS level crossing, having the same four total momenta 
${\bf K}=2\pi/6(\pm 1,\pm 1)$ for the four GSs when $t \rightarrow 0$ 
and when $U \rightarrow 0$.

This absence of GS level crossing is a property of the case $N=L/2$. 
If $L/2 > N=3$, the GS momentum is ${\bf K}=0$ when $t \rightarrow 0$, 
and a GS level crossing takes place as $r_s$ increases. This correlated 
lattice regime in the limit $L \rightarrow \infty$ is discussed in 
App. \ref{Appendix B}. 

 Let us now consider one of the GS wave functions of momentum 
$\bf{K}$, for instance ${\bf K}=2\pi/6 (1,1)$. When $t \rightarrow 0$, 
it reads  
\begin{equation}
\left|\Psi_0({\bf K})\right>= {1\over 6} \sum_{\bf j} e^{i{\bf Kj}} 
D^\dagger_{\bf j} \left|0\right>, 
\end{equation}  
and corrections are given at the leading order (first order) 
of the $t/U$ expansion by:
\begin{equation}
\left|\Psi_0^{(I)}({\bf K})\right>=\left|\Psi_0({\bf K}) \right>+ 
\sum_{\alpha \atop \alpha\neq 0}
{\left<\Psi_{\alpha}({\bf K})\left|H_1\right|\Psi_0({\bf K})\right>
\over E_0^{(0)}-E_{\alpha}}\left|\Psi_{\alpha}\right>
\label{pertseries}
\end{equation}
Only twelve $t \rightarrow 0$ eigenstates $\left|\Psi_{\alpha'}({\bf K}) 
\right>$ are directly coupled to  $\left|\Psi_0({\bf K})\right>$ by a 
single hop (coupling term $-t$). This gives: 
\begin{equation}
\left|\Psi_0^{(I)}({\bf K})\right>=\left|\Psi_0({\bf K})\right>+
\sum_{\alpha'=1}^{12}{t\over \Delta E_{\alpha'}}\left|\Psi_{\alpha'} 
({\bf K}) \right>,
\end{equation}
where $\Delta E_{\alpha'} = E_{\alpha'}-E_0^{(0)}$.
The $t/U$ expansion of the average 
$\left<\Psi_0({\bf K},r_s)\left|f\right|\Psi_0({\bf K},r_s)\right>$ 
of an observable $f$ which takes definite values 
over the states $\left| \Psi_{\alpha'} ({\bf K})\right>$ are given 
at leading order ($\propto (t/U)^2$) by:
\begin{equation}
\left<f\right>^{(II)}=\left({1\over C}\right)^{(II)} f(\Psi_0)+
\sum_{\alpha'=1}^{12}\left({t\over\Delta E_{\alpha'}}\right)^2 
f(\Psi_{\alpha'}),
\end{equation}
where 
\begin{equation}
\left({1\over C}\right)^{(II)} = 1- \sum_{\alpha'=1}^{12}
\left({t\over\Delta E_{\alpha'}}\right)^2.
\end{equation}

\begin{figure}[b]
{\centerline{\leavevmode \epsfxsize=9cm \epsffile{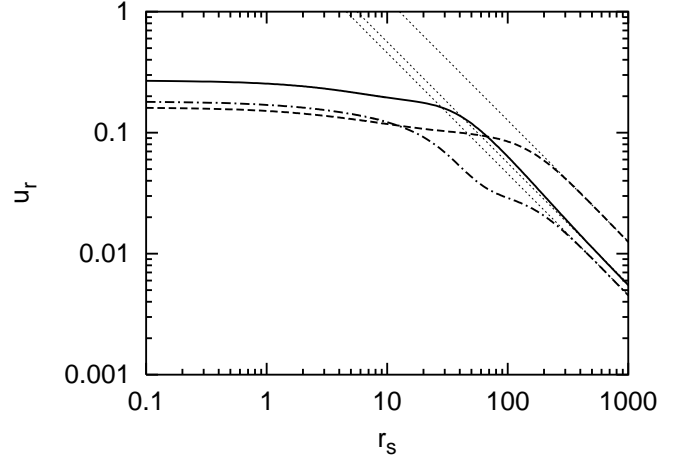}}}	
\caption{The relative fluctuations $u_r$ of the three inter-particle 
spacings $r_{min}$ (solid line) $r_{int}$ (dashed-dotted line) and 
$r_{max}$ (dashed line) as a function of $r_s$. The thin dotted 
lines give the perturbative $t/U$-decays given after Eq. \ref{u_r3}.}
\label{fig42}
\end{figure}

 This can be used to calculate the $n^{th}$ moment of the 
different inter-particle spacings in the correlated lattice 
limit. Each state 
\begin{equation}
\left|ijl\right>=c^{\dagger}_{\vec{i}}c^{\dagger}_{\vec{j}}
c^{\dagger}_{\vec{l}} \left|0\right>
\end{equation}
is characterized by three inter-particle spacings 
$r_{min} (\vec{ijl}) \leq r_{int}(\vec{ijl}) \leq 
r_{max}(\vec{ijl})$. Taking for $f$ 
\begin{equation}
f(r^n)=\sum_{\bf ijl} \left|\bf{ijl}\right> 
r^n (\bf{ijl}) \left<\bf{ijl}\right|
\end{equation}
one can calculate the averages $\left<r\right>$, the variances 
$\left<r^2\right>-\left<r\right>^2$ and the relative fluctuations 
\begin{equation} 
u_r=\sqrt{\frac{\left<r^2\right>}{\left<r\right>^2}-1}
\label{u_r3}
\end{equation}
of $r_{min}$, $r_{int}$ and $r_{max}$ in the correlated lattice 
limit. 

 The three relative fluctuations $u_r$ are shown in  Fig. \ref{fig42}, 
as a function of $r_s$. One can see that above $r_s^*$, they 
coincide with the behaviors $u_r(r_{min}) \approx 5.49t/U$, 
$u_r(r_{int}) \approx 4.47t/U$ and $u_r(r_{max}) \approx 12.34t/U$ 
given by the leading order of the $t/U$ expansion.

\section{Breakdowns of the perturbative expansions}
\label{section5}

\begin{figure}[b]
{\centerline{\leavevmode \epsfxsize=9cm \epsffile{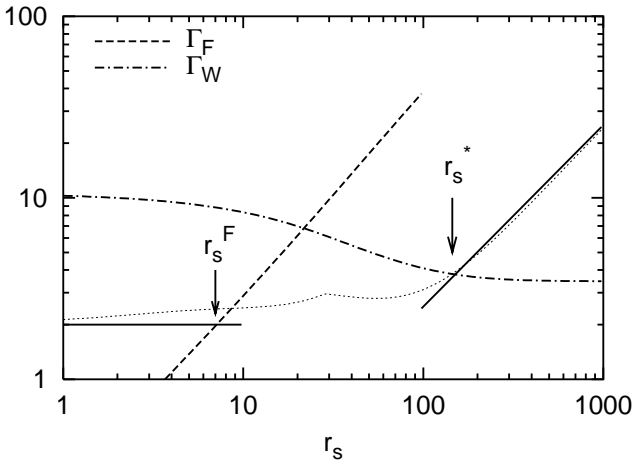}}}
\caption
{ First energy spacing $\Delta E (r_s)$ (dotted line) with its  
two limits (solid lines - $2t$ for $U \rightarrow 0$ and $0.025 U$ 
for $t \rightarrow 0$), and spreading widths $\Gamma_F(r_s)$ 
(dashed line) and $\Gamma_W (r_s)$ (dashed-dotted line) calculated from  
the local densities of states $\varrho_F(E)$ and $\varrho_W(E)$ 
respectively. $L=6$ and ${\bf K}=(2\pi/6,2\pi/6)$. The arrows indicate the limits 
$r_s^F$ and $r_s^*$ between which the perturbative expansions are 
inappropriate.
} 
\label{figb1}
\end{figure}

 When a perturbation is weak, the eigenstates remain localized in the 
vicinity of the unperturbed states. Increasing the perturbation 
delocalizes the eigenstates in the unperturbed eigenbasis, yielding 
a crossover \cite{wp} from a weak perturbative mixing of the unperturbed 
eigenstates (Rabi oscillations) towards an effective golden-rule decay. 
Above the delocalization threshold, an expansion around the unperturbed 
eigenbasis does not make sense. This delocalization threshold is given by 
a general criterion discussed in different contexts: onset of quantum chaos 
in a many body spectrum \cite{wp,shepelyansky-suskhov,wpi}, quasi-particle 
lifetime and delocalization in Fock space \cite{altshuler,jacquod}), 
interaction induced thermalization \cite{flambaum-casati}: Quantum ergodicity 
occurs when the perturbation matrix element $\left<i|P|f\right>$ between an 
unperturbed eigenstate $\left|i\right>$ to the ``first generation'' of 
unperturbed eigenstates $\left|f\right>$ directly coupled to it by the 
perturbation is of the order of their level spacing $E_f-E_i$: 
\begin{equation}
\left<i|P|f\right> \approx E_i-E_f. 
\label{deloc}
\end{equation}
 Using this general criterion for the three particle GS, one can 
estimate the range of intermediate ratios $r_s$ where the GS cannot 
be simply described neither in terms of the low energy weak coupling 
eigenstates, nor in terms of the low energy strong coupling eigenstates.

\subsection{Limit of the weak coupling $U/t$-expansion}

\begin{figure}
{\centerline{\leavevmode \epsfxsize=9cm \epsffile{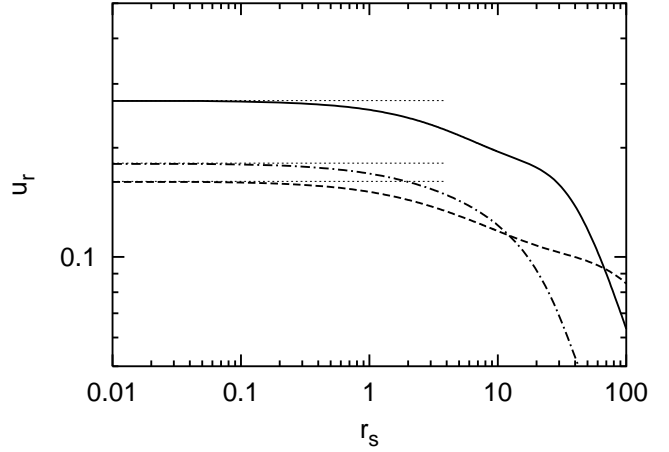}}}	
\caption{The relative fluctuations $u_r$ of the three distances 
$r_{min}$, $r_{int}$ and $r_{max}$ as a function of $r_s$. The 
dotted lines give the HF values, which coincide to the actual 
behaviors up to $r_s^{HF} \sim 1$.}
\label{fig32}
\end{figure}

\begin{figure}
{\centerline{\leavevmode \epsfxsize=9cm \epsffile{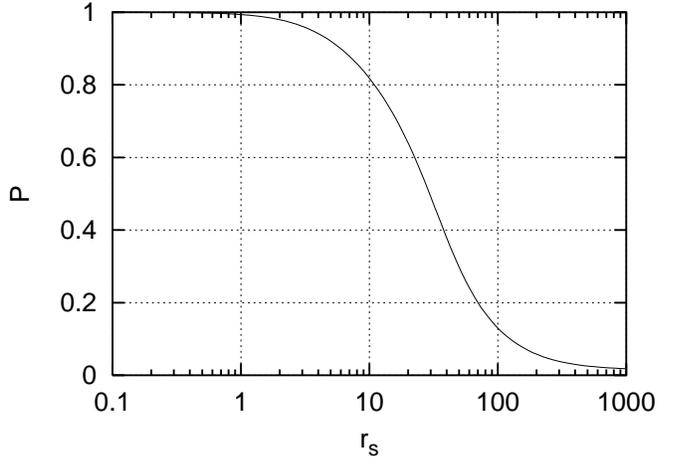}}}	
\caption{The GS projection $|\left<\Psi_0(r_s)|\Psi_0(r_s=0)\right>|^2$ 
over the non-interacting GS (or HF-GS) as a function of $r_s$. 
Notice the breakdown of the HF-behavior at $r_s^{HF} \sim 1$.}
\label{fig33}
\end{figure}

 When $U/t \rightarrow 0$, one can use the Hartree-Fock (HF) 
mean-field approximation for the ground state. Since the charge 
distribution is  uniform without disorder and with periodic BCs, 
this approximation becomes trivial. Starting from  plane wave 
states of uniform density, the HF-approximation consists in 
studying again a single particle in a uniform potential. Therefore, 
the GSs for $U=0$ remain the self-consistent states of the 
HF-approximation, while the HF-eigenenergies are given by the first 
order pertubative expressions (see Eqs. \ref{U-expansion1} 
and \ref{U-expansion2}). The 
range of validity of the HF-approximation can be seen in 
Fig. \ref{fig31}, Fig. \ref{fig32} and Fig. \ref{fig33} 
for various observables. As one varies $r_s$ and for $L=6$, 
Fig. \ref{fig31} gives the GS-energies, for the total momenta 
$\vec{K}=2\pi/L (1,1)$ and $\vec{K}=2\pi/L (0,0)$ respectively. 
The relative fluctuations $u_r(U)$ of the three inter-particle 
spacings and the projection 
\begin{equation}
P_{HF}(U)=\left|\left<\Psi_0(U=0)\mid \Psi_0(U)\right>\right|^2
\end{equation}
of the actual GSs onto the HF-GSs are given in Fig. \ref{fig32} and 
Fig. \ref{fig33} respectively, for a total momentum $\vec{K}=2\pi/6 (1,1)$. 
These figures show that the HF-approximation remains valid up to a value 
$r_s^{HF} \approx 1$. 

\begin{figure}
{\centerline{\leavevmode \epsfxsize=9cm \epsffile{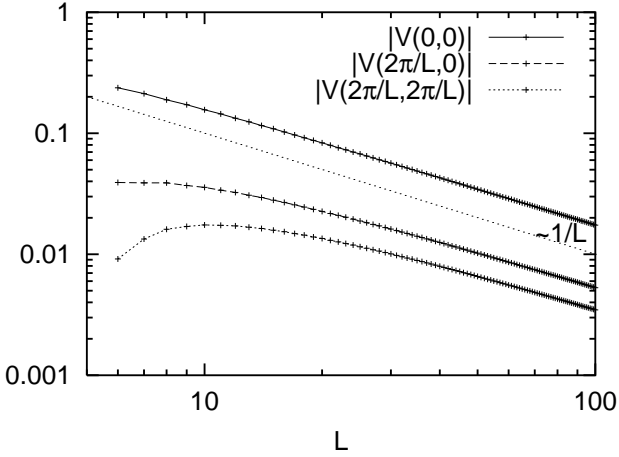}}}	
\caption{The dimensionless interaction matrix elements 
$|V({\bf q})|$ (see Eq. \ref{int}) as a function of $L$ for 
different values of ${\bf q}$, exhibiting the asymptotic $1/L$ 
decay.}
\label{fig-interaction}
\end{figure}

 This numerical value for $r_s^{HF}$ is close to the value $r_s^{F}$ 
yielded by the general criterion (Eq. \ref{deloc}) giving the 
crossover from a weak perturbative mixing of the unperturbed states 
towards an effective golden-rule decay (delocalization threshold in 
Fock space). 

Let us first give a qualitative estimate valid for arbitrary values 
of $L$ and $N$, where the values of $U$, $t$, $N$ and $L$ appear 
through the expected dimensionless ratio $r_s$. Assuming $L$ 
sufficiently large for having $\cos(2\pi/L)\approx 1-2 (\pi/L)^2$, 
the energy spacing $\Delta E=E_1-E_0$ between the GS and the first 
excitations for $U=0$ reads: 
\begin{equation}
\Delta E = 2\pi t \sum_i \left({\bf k}_{i1}^2-{\bf k}_{i0}^2\right) 
\approx 2 \pi t \left(2 {\bf k}_F {\bf q}\right) \sim {t \sqrt{N}\over L^2},
\end{equation}
since ${\bf k}_F \sim \sqrt{N}/L$ and ${\bf q}\sim 1/L$. The interaction 
matrix element directly coupling these states reads:
\begin{equation}
\left< 0 |H_{int}({\bf q})| 1\right> \sim {U\over L}, 
\end{equation}
as shown in Fig. \ref{fig-interaction} for a large enough $L$.
The criterion (Eq. \ref{deloc}) 
$\Delta E \approx \left< 0 |H_{int}({\bf q})| 1\right>$ 
gives:
\begin{equation}
{U^F L \over t \sqrt{N}} = const. \rightarrow r_s^F = const.
\end{equation}

\begin{figure}[t]
{\centerline{\leavevmode \epsfxsize=9cm \epsffile{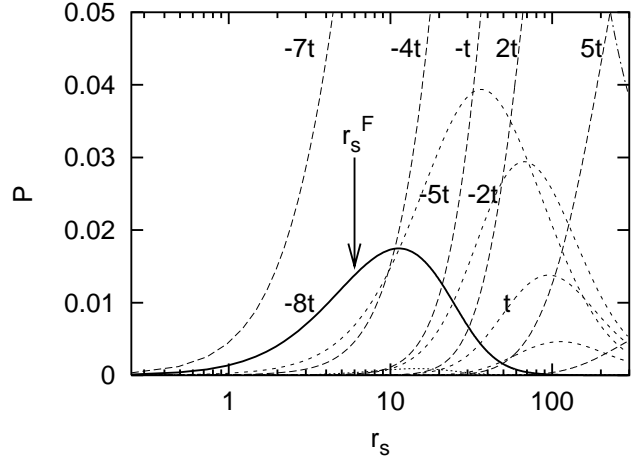}}}
\caption {GS projection $|\left<\Psi_0(r_s)|\Psi_{\alpha}(r_s=0)\right>|^2$ 
onto the non-interacting eigenspaces of increasing energies 
$E_{\alpha}= -8t,-7t,\ldots$ and total momentum ${\bf K}=(2\pi/L,2\pi/L)$ 
as a function of $r_s$. Notice the behaviors of the projections onto the 
$\left|\Psi_{\alpha}(r_s=0)\right>$ which do not contribute to 
$\left |\Psi_0(r_s \rightarrow \infty)\right>$ 
($E_{\alpha}=-8t,-5t,-2t,t, \ldots$).
}
\label{supe1}
\end{figure}

 For an exact  determination of $r_s^F$ on $6\times6$ 
lattice, we have numerically studied the local density of states 
(LDOS) \cite{flambaum-casati},
\begin{equation}
\varrho_F(E)=\sum_i \bigl|\left<\Psi_0(U=0)\mid \Psi_i(U) \right>\bigr|^2
\delta(E-E_i(U))
\end{equation}
of the non-interacting GS in the eigenbasis with interaction. This 
LDOS is a distribution of width $\Gamma_F$, where $\hbar/\Gamma_F$ 
gives the lifetime of the $U=0$ states when $U$ is turned on. 
Using Fermi golden-rule for estimating $\Gamma_F$:
\begin{equation}
\Gamma_F = 2\pi \sum_1 |\left<0|H_{int}|1\right>|^2 \varrho(E_1)
\sim U^2 \left|V({\bf q})\right|^2 \frac{1}{E_1-E_0}, 
\end{equation}
and since the density of states $\varrho(E_1) \approx 1/\Delta E$, 
criterion (\ref{deloc}) corresponds to 
\begin{equation}
\Gamma_F \approx \Delta E. 
\end{equation}
For ${\bf K}=2\pi/6,2\pi/6)$, the GS of energy $-10t$ is directly 
coupled to two states of energy $-8t$ by a matrix element 
$2U (V(2\pi/6,2\pi/6)-V(2\pi/6,0))= 0.091 U$, and the condition  
\begin{equation}
\Gamma_F= 4\pi {(0.091 U)^2 \over 2t} \approx \Delta E=2t,  
\end{equation}
is satisfied when $r_s^F \approx 6$. 
We have also calculated $\Gamma_F$ directly from 
$\varrho_F(E)$, and one obtains again $\Gamma=\Delta E$ at 
$r_s^F \approx 6$, as shown in Fig. \ref{figb1}. One can 
notice that 

\begin{itemize} 

\item The ratio $r_s^F$ (breakdown of a $U/t$ expansion) 
is larger than the ratio $r_s^{HF} \approx 1$, where the HF-behaviors 
cease to be valid (zero order approximation for the wave function, 
first order approximation for the energies). 

\item Fig. \ref{supe1} gives the GS projection over the non-interacting 
eigenspaces of increasing energies. The projection onto the 
first excitations $\left |1 \right >$ of energy $-8t$ is particularly 
interesting. These states are directly coupled by the interaction to 
the GS for $U=0$, but are orthogonal to the GS when $ U \rightarrow 
\infty $. One can see that the increase of this projection driven 
by the $U/t$ expansion ceases precisely at $r_s^F$. This is the first 
manifestation of the correlated limit as $U$ increases.  

\end{itemize} 

\subsection{Limit of the $t/U$ correlated lattice expansion}

 We use the criterion (Eq. \ref{deloc}) for determining the value 
$r_s^*$ under which the correlated lattice expansion breaks down.
The matrix element $|\left<0|H_{kin}|1 \right>|=t$ 
is equal to the first corresponding energy spacing $\Delta E$ 
for $U=U^*$, where $U^*$ satisfies 
\begin{equation}
t = \Delta E \approx \sqrt{2} \frac{U^*}{L^2} + O (L^{-3}),
\end{equation}
which yields the threshold: 
\begin{equation} 
{U^*\over tL^2} \approx const \rightarrow r_s^* \approx const \cdot L^3. 
\end{equation}
When $r_s < r_s^* \propto L^3$, the GS becomes delocalized in the 
$r_s \rightarrow \infty$ eigenbasis and the $t/U$ expansion breaks down.
This gives the same $L$-dependence for $r_s^*$ than in 
Sec. \ref{section2}. 

From the local density of states
\begin{equation}
\varrho_W(E)=\sum_i \bigl|\left<\Psi_0(U=\infty)\mid \Psi_i(U)
\right>\bigr|^2 \delta(E-E_i),
\end{equation}
of the GS $\left|\Psi_0(U=\infty)\right>$ when  $U\rightarrow \infty$ 
in the eigenbasis with interaction, we have calculated the spreading width 
$\Gamma_W$ of $\left|\Psi_0(U=\infty)\right>$ when one turns on the kinetic 
hopping term $t$. One can see in Fig. \ref{figb1} that $\Gamma_W=\Delta E$ 
for $r_s^* \approx 180$. Above $r_s^*$, the three inter-particle GS spacings 
are correctly given by the $t/U$-expansion (see Fig. \ref{fig42}) and 
the ordering of the $L^2$ low energy levels correspond to those given 
in Table \ref{pertres1}, the last level crossing being shown 
in Fig. \ref{fig41} at $r_s \approx r_s^*$.  
  
\section{Partially melted Wigner molecule near the correlated lattice limit}
\label{section6}

We begin to study the GS in the non-perturbative regime $r_s^F<r_s<r_s^*$. 
We first introduce a simple ansatz proposed in Ref. \cite{nemeth} and 
which turns out to describe the GS for $r_s \approx 40$. The idea of this 
ansatz can be given by three observations:

\begin{figure}[b]
{\centerline{\leavevmode \epsfxsize=4cm \epsffile{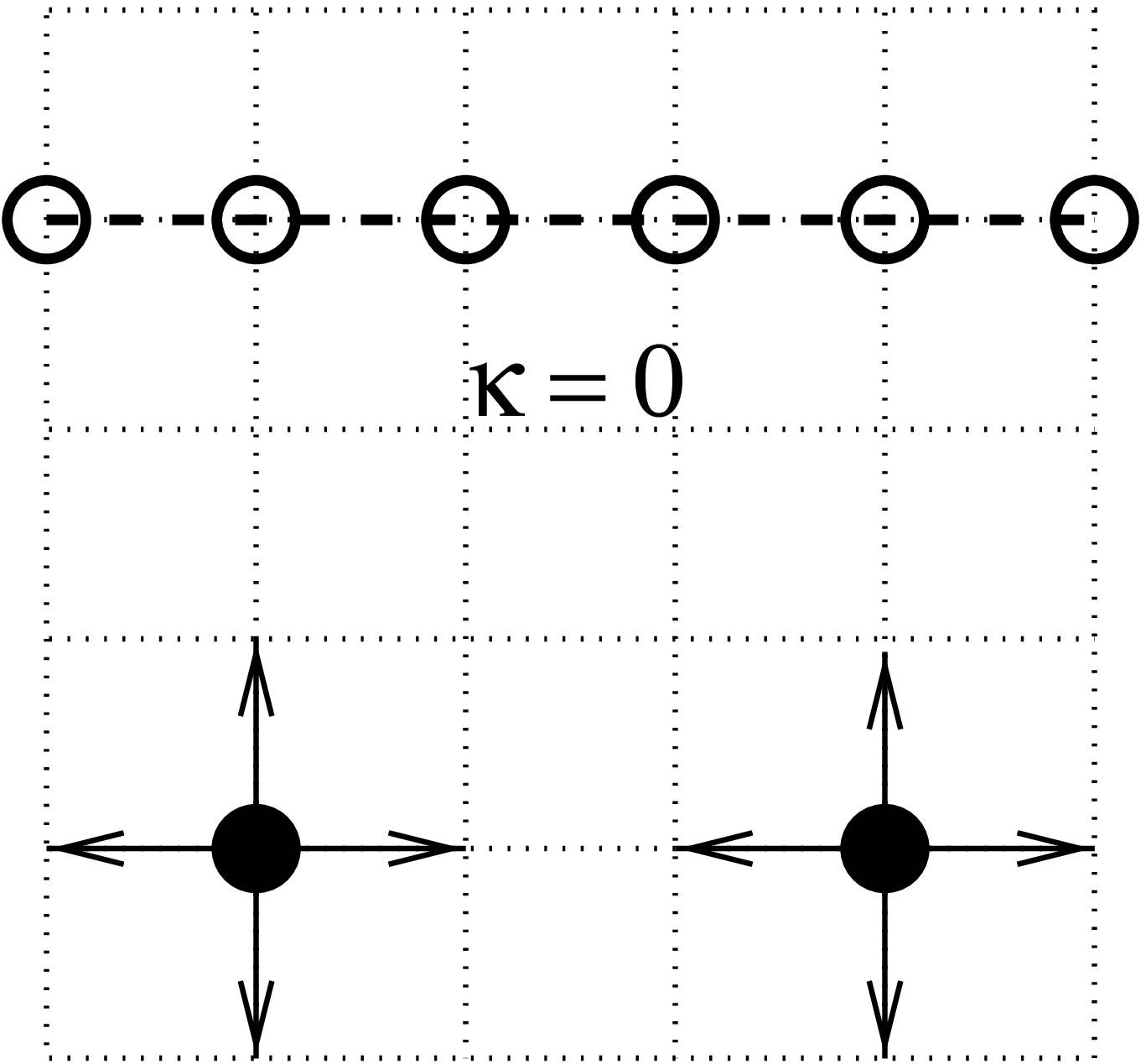}}}
{\centerline{\leavevmode \epsfxsize=9cm \epsffile{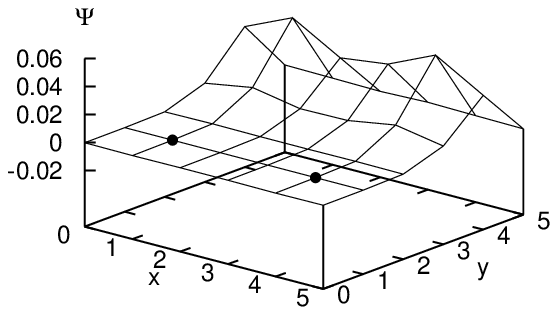}}}	
\caption{{\bf Above}: Scheme of a $x$-oriented PMWM: one
particle is totally delocalized in $x$-direction with a momentum 
$\kappa$, but remains localized in the $y$-direction at a distance 
$L/2=3$ from the two other particles which form a 2PWM.
{\bf Below}: The GS projection amplitude $\psi(x,y)$ (see Eq. \ref{eq-proj}) 
at $r_s=40$. $\psi(x,y)$ is not totally uniform in the $x$-direction, 
since there is the contribution of the $y$-oriented symmetric PMWM.}
\label{fig51}
\end{figure}

\begin{itemize}

\item As shown in Fig. \ref{fig401} and Fig. \ref{fig40}, when $r_s$ is 
large, the three particles form a triangle, which can be seen as a square 
with a `vacancy' at one of its corners. Therefore, beside the rigid 
translation of the triangle (hopping term $h \propto t^3/U^2$), there is 
another possible third order process: the tunneling of the vacancy 
(hopping term $r \propto t^3/U^2$). As $r_s$ decreases, the GS may have 
advantage to delocalize this vacancy to reduce its kinetic energy. This 
will be the ``Andreev-Lifshitz supersolid'' in an ultimate 
mesoscopic limit.

\item If one looks at Fig. \ref{fig42}, we see that the smallest 
inter-particle spacing $r_{min}$ begins to fluctuate according to 
the lattice $t/U$-expansion above $r_s \approx 40$ while the largest 
one $r_{max}$ enters into this lattice regime at a higher value $r_s \approx 
r_s^* \approx 180$. The behaviors of $r_{min}$ and $r_{max}$ suggest us that 
one has for $ 40 < r_s <180$ a partially melted Wigner molecule (PMWM). 
This PMWM is composed of a rigid two particle Wigner molecule (2PWM) with 
oscillations around the equilibrium positions smaller than the lattice 
spacing $a$ (consistent with the correlated lattice behavior of $r_{min}$), 
while the third particle remains more delocalized, explaining the 
absence of a correlated lattice behavior of $r_{max}$. 
 
\item One can see the delocalization of the third particle 
in the direction parallel to the 2PWM formed by the two others, 
in Fig. \ref{fig51}, where the GS projection amplitude 
\begin{equation}
\psi(x,y)=\left<\Psi_0(r_s)\left|c^{\dagger}_{(1,1)}c^{\dagger}_{(4,1)}
c^{\dagger}_{(x,y)}\right|0\right>
\label{eq-proj}
\end{equation}
is given for $r_s=40$. 
\end{itemize} 

 Let us consider a x-oriented state where two particles are fixed, while 
the third particle is delocalized along one line with a momentum $\kappa$, 
as shown in Fig. \ref{fig51}:
\begin{equation}
c_{\bf j}^\dagger c_{{\bf j}+(3,0)}^\dagger c_{\kappa,j_y+3}^\dagger  
\left|0\right>.
\end{equation}
To satisfy translational symmetry, one defines the $x$-oriented PMWM 
of total momentum ${\bf K}$ by 
\begin{eqnarray}
\left|\Psi^x_\kappa({\bf K})\right>&=& {1\over N} \sum_{\bf j} 
e^{i ({\bf K j}-\kappa j_x)} \nonumber \\
&&\cdot \sum_{j_x'} e^{i \kappa j_x'} 
c_{\bf j}^\dagger c_{{\bf j}+(3,0)}^\dagger
c_{(j_x',j_y+3)}^\dagger \left|0\right>.
\end{eqnarray}
Then, we combine the $x$-oriented  PMWM with its symmetric $y$-oriented 
counterpart:
\begin{equation}
\left |\Psi_\kappa({\bf K}) \right>=\sqrt{3\over8} 
\left(\Psi^x_\kappa({\bf K})+\Psi^y_\kappa({\bf K})\right)
\end{equation}
for satisfying the $x\leftrightarrow y$ symmetry. Moreover, $\kappa$ 
can only take three quantized values ($0, 4\pi/6, 8\pi/6$) for $L=6$.

\begin{figure}[t]
{\centerline{\leavevmode \epsfxsize=9cm \epsffile{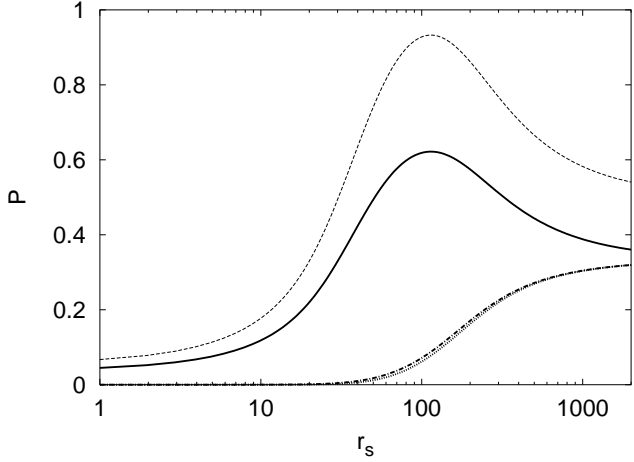}}}	
\caption
{The GS-projections for ${\bf K}=2\pi/6(1,1)$ over different $x$-oriented 
PMWMs $\Psi_\kappa^x({\bf K})$: solid ($\kappa=0$), dash-dotted 
($\kappa=4 \pi/6$) dotted ($\kappa=8\pi/6$) lines respectively. 
The upper thin dashed line gives the total GS-projection over 
$x\leftrightarrow y$ symmetric $\left|\Psi_{\kappa=0}({\bf K})\right>$ 
}
\label{fig52}
\end{figure}

\begin{figure}
{\centerline{\leavevmode \epsfxsize=9cm \epsffile{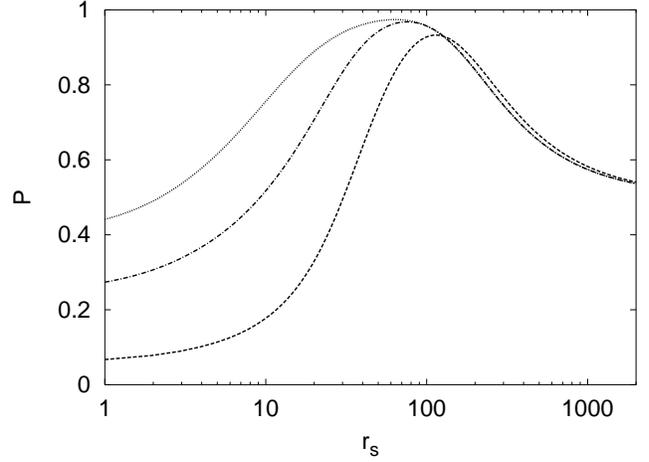}}}	
\caption{GS-projection over the PMWM-ansatz $\tilde\Psi_{\kappa=0}({\bf K})$ 
when the oscillatory motions of the PMWMs are included up to 0th (dashed), 
1st (dashed-dotted) and 2nd (dotted) orders of a lattice $t/U$-expansion. 
}
\label{fig53}
\end{figure}

\begin{figure}
{\centerline{\leavevmode \epsfxsize=9cm \epsffile{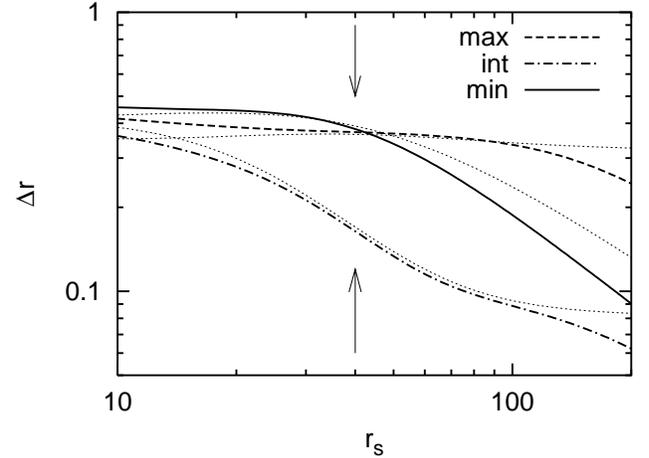}}}	
\caption{Root mean square $\Delta r$ of the three inter-particle spacings 
$r_{min}$, $r_{int}$ and $r_{max}$ as a function of $r_s$: actual behaviors 
(thick lines) and behaviors given by the 2nd order PMWM-ansatz (dotted 
lines). The arrows underline the close agreement for $r_s\approx 40$. 
}
\label{fig54}
\end{figure}

 One can follow the delocalization of the third particle in Fig. \ref{fig52}. 
When $r_s \rightarrow \infty$, the GS-projections 
$\left|\left<\Psi_\kappa^x({\bf K})\mid \Psi_0(r_s)\right>\right|^2$ over 
the $x$-oriented PMWM states of $\kappa = 0, 4\pi/6, 8\pi/6$ 
become equal. This corresponds to a total localization of the third 
particle on the line parallel to the 2PWM, such that the three 
particles form rigid triangles. When $r_s \leq 40$, only the contribution 
with $\kappa = 0$ is non-zero: the third particle becomes fully delocalized. 
Moreover, if we take the $x\leftrightarrow y$ symmetric combination, we can 
see that we describe more than 90 \% of the the real GS when  
$r_s\approx 100 < r_s^*$. If $r_s$ is further decreases, our ansatz have to 
include some oscillatory motions of the 2PWMs around equilibrium. 
If these oscillations are smaller than the lattice spacing, it is enough to 
use a similar lattice $t/U$-expansion as in (\ref{pertseries}),  for the 
2PWMs only. Fig. \ref{fig53} shows the improvement of the GS-projection 
$\left|\left<{\tilde \Psi}_\kappa({\bf K})\mid \Psi_0(U)\right>\right|^2$
when the PMWM oscillations are included. The GS-projection over the expanded 
ansatz exceeds now 95 \% at a smaller value of $r_s \approx 40$ if the 
$t/U$-expansion is extended up to the 2nd order. 

 Fig. \ref{fig54} shows how the PMWM-ansatz, with 2nd order oscillatory 
motions of the 2PWMs, describes the root mean square of the distribution 
of the three inter-particle spacings $r_{min}$, $r_{int}$ and $r_{max}$ 
around $r_s \approx 40$. 

 Above $r_s^* \approx 180$, the ``triangles'' are formed. Below $r_s^*$, 
the gradual delocalization of one particle parallel to a remaining 2PWM 
with increasing oscillations begins. Around $r_s \approx 40$, this 
delocalization is completed with a momentum $\kappa=0$. This gives a mixed 
state, having both the properties of a ``solid'' and of a ``liquid''. 
However this mixed state exhibits lattice effects and cannot be the 
final step of the melting process. 

\section{Partially melted Wigner molecule near the Fermi limit}
\label{section7}

When the interaction is further decreased, two things have to be taken 
into account. 

\begin{itemize}

\item One concerns the ``solid'': the amplitudes of the oscillatory 
motions begin to exceed the lattice spacing, and cannot be described 
by the lattice $t/U$ expansion. 

\item The other concerns the ``liquid'': the Coulomb repulsion cannot 
be strong enough to maintain the delocalized particle parallel to the 
correlated pair. 

\end{itemize}

To describe the next step of melting, in the range $r_s^F < r_s < 40$, let 
us consider the Fermi limit and the picture \cite{bouchaud} suggested by 
Bouchaud et al of a system of unpaired particles with a reduced Fermi 
energy co-existing with strongly paired, nearly solid assembly. For the 
ultimate limit $N=3$, this simply means a single correlated pair 
co-existing with a third particle remaining in its non-interacting 
state. The third particle roughly remains in a plane wave state of 
uniform density (momentum ${\bf k}_3 = (0,0)$, $2\pi/6(1,0)$ or 
$2\pi/6(0,1)$ if ${\bf K}=2\pi/6(1,1)$), and provides a uniform background 
for the correlated pair formed by the two other particles. The occurrence 
of a correlated pair is suggested by the breakdown of the $U/t$-expansion 
above $r_s^F$ on one side, by the previous PMWM ansatz on the other side. 
The conservation of the total momentum allows us to order these 2+1 particle 
systems. Different combinations, given in Table \ref{2pstates}, are possible.
\begin{table}[hbt] 
\begin{center}
\begin{tabular}{||c||c|c|c||} \hline
$\Psi_\alpha$&${\bf k}_{3\alpha}$  &${\bf K}_{2\alpha}$  &$\bf a_\alpha$ \\ \hline \hline
$\Psi_1$&$2\pi/6(0,1)$&$2\pi/6(1,0)$&$(3,3)$ \\ \hline
$\Psi_2$&$2\pi/6(1,0)$&$2\pi/6(0,1)$&$(3,3)$ \\ \hline
$\Psi_3$&$2\pi/6(0,0)$&$2\pi/6(1,1)$&$(3,2)$ \\ \hline
$\Psi_4$&$2\pi/6(0,0)$&$2\pi/6(1,1)$&$(2,3)$ \\ \hline
\end{tabular}
\end{center}
\caption{Four different combinations of an unpaired particle of momentum 
${\bf k}_{3\alpha}$ 
co-existing with a correlated pair of momentum ${\bf K}_{2\alpha}$.  
${\bf K}_{2\alpha}+{\bf k}_{3\alpha}={\bf K}=2\pi/6(1,1)$ and $\bf a_\alpha$ 
is the pair inter-particle spacing when $U \rightarrow \infty$.
\label{2pstates}}
\end{table}

 In this table, $\bf a_\alpha$ is the asymptotic value of the inter-particle 
spacing of the pair, if we follow the level to the limit 
$U \rightarrow \infty$, the asymptotic pair wave function 
becoming $c^\dagger_{\bf j} c^\dagger_{{\bf j}+{\bf a_\alpha}} 
\left|0\right>$. 
However, for ${\bf K}_2=2\pi/6(1,1)$, this pair wave function with 
${\bf a} = (3,3)$ is zero. In this case, the pair GS has a twofold 
degeneracy, with ${\bf a} = (3,2)$ and ${\bf a} = (2,3)$. This 
is why we have four combinations instead of three, when we keep 
one particle out of three in its non-interacting wave function.

\begin{figure}
{\centerline{\leavevmode \epsfxsize=9cm \epsffile{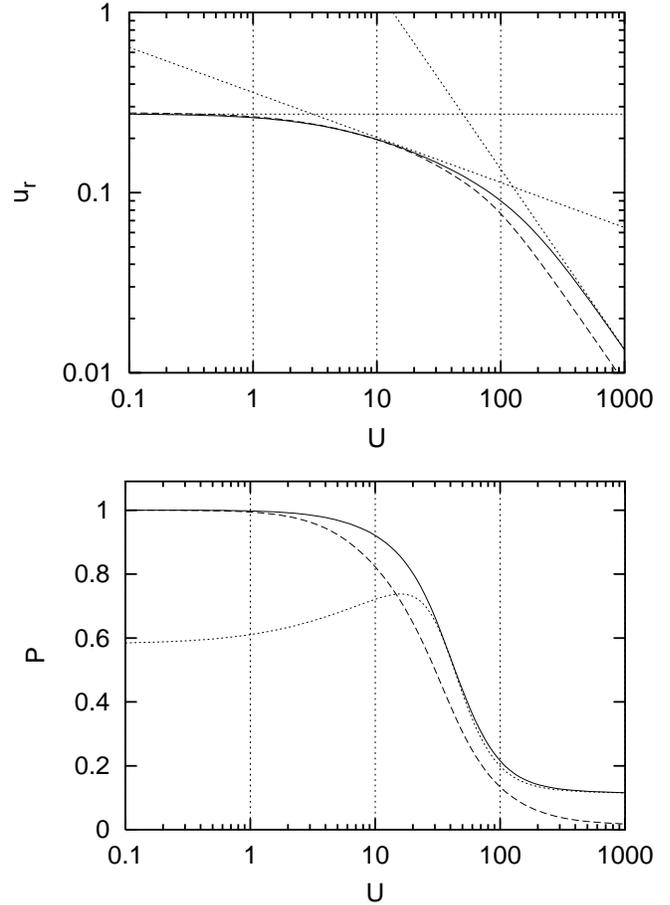}}}	
\caption
{
{\bf Above}: $N=2$ particles: Relative fluctuations $u_r(U)$ of the 
pair GS (solid line - asymptotic spacing ${\bf a}=(3,3)$) and first 
excited state (dashed line -  ${\bf a}=(3,2)$ or $(2,3)$). The dotted 
lines give the three regimes for the pair: (Fermi liquid - continuous Wigner 
solid - correlated lattice solid).{\bf Below}: $N=3$ particles: GS-projection 
$P_0(U)$ (dashed line) on the non-interacting GS, and GS-projection 
($P_A(U)$, solid line) and ($P_A^{(II)}(U)$, dotted line) onto the 
four states given in table \ref{2pstates} - the pair wave function being 
calculated exactly or using the 2nd order $t/U$-expansion respectively.
}
\label{fig61} 
\end{figure}

 Let us begin to consider those four states $\left|\Psi_{\alpha}\right>$, 
assuming that the pair behaves as rigidly as in the limit 
$r_s \rightarrow \infty$:
\begin{equation}
\left|\Psi_\alpha\right>=\sum_{\bf j} e^{i{\bf K}_{2\alpha}{\bf j}} 
c^\dagger_{\bf j} c^\dagger_{{\bf j}+{\bf a_\alpha}} 
c^\dagger_{{\bf k}_{3\alpha}}\left|0\right>.
\end{equation}
Let us first allow oscillatory motion of the pair using the 2nd order 
lattice $t/U$-expansion. In Fig. \ref{fig61}, one can see the 
total GS-projection
\begin{equation}
P_A^{(II)}(U)=\sum_{\alpha=1}^4\left|\left<{\Psi}^{(II)}_\alpha
\mid\Psi_0(U)\right>\right|^2
\end{equation}
onto these four states. As $U$ decreases, $P_A^{(II)}(U)$  increases, but 
saturates below $U \approx 20$. This saturation tells us that the lattice 
$t/U$ perturbation theory is no longer a suitable tool for describing the 
pair. To improve the GS description, we numerically calculate the exact 
wave function $\left| \Phi_{{\bf K}_{2\alpha}}(U) \right> $ of the pairs: 
\begin{equation}
\left| \Phi_{{\bf K}_{2\alpha}}(U) \right> = \sum_{{\bf k}_1,{\bf k}_2 
\atop {\bf k}_1+{\bf k}_2={\bf K}_{2\alpha}} 
\Phi^\alpha_{{\bf k}_1,{\bf k}_2}(U) c^\dagger_{{\bf k}_1} 
c^\dagger_{{\bf k}_2}\left|0\right>;
\end{equation}
to have the following ansatz wave function for each combinations:
\begin{equation}
\left|\bar\Psi_\alpha(U)\right> = \left( \sum_{{\bf k}_1,{\bf k}_2 \atop 
{\bf k}_1+{\bf k}_2={\bf K}_{2\alpha}} \Phi^\alpha_{{\bf k}_1,{\bf k}_2}(U)
c^\dagger_{{\bf k}_1} c^\dagger_{{\bf k}_2} \right) 
c^\dagger_{{\bf k}_{3\alpha}}\left|0\right>.
\end{equation}
The total GS-projection 
\begin{equation}
P_A(U)=\sum_{(\alpha)}\left|\left<{\bar\Psi}_\alpha(U) 
\mid\Psi_0(U)\right>\right|^2
\end{equation}
over the subspace spanned by the four ansatz states is given by a 
solid line in Fig. \ref{fig61} 
(below). For comparison, we have also plotted the GS-projection 
\begin{equation}
P_0(U)=\left|\left<\Psi_0(U=0) \mid\Psi_0(U)\right>\right|^2
\end{equation}
onto the the non-interacting GS (dashed line) of same $\bf K$. 
If one uses  unpaired fermions co-existing with correlated 
pairs instead of the non-interacting GS, one substantially  
improves the description of the intermediate GS above $r_s^F$. 
For $U\approx r_s =10$, the projection $P_A(U)$ exceeds 90 \%. 
In Fig. \ref{fig61}, one can see for the same value of $U$ the 
relative fluctuation $u_r$ of the corresponding two particle 
system, both for the ${\bf a}_\alpha=(3,3)$ and $(3,2)$-states. 
This study of the case $N=2$ shows that the pairs are neither in their 
solid lattice regime, nor in their liquid non-interacting limit 
when $U\approx 10$, but in their continuous Wigner regime.

\section{Andreev-Lifshitz supersolid for intermediate couplings?}
\label{section9}

 With this second ansatz, our understanding of the different steps 
of the melting of the three particle system on a $6 \times 6$ lattice 
is achieved. Above $r^*_s\approx 180$, we have delocalized correlated 
triangles and a lattice $t/U$ expansion is sufficient for  
describing the $L^2$ low energy states. For the GS, the melting begins 
with the partial delocalization of one particle in the $x$ or $y$ 
directions down to $r_s\approx 40$. This process is the natural 
extension of the vacancy tunneling which is the ``softest'' degree 
of freedom of the correlated triangle, if one views a triangle 
as a square with a vacancy. This delocalization of a single particle 
is first one dimensional, parallel to the remaining 2PWMs. For $10<r_s<40$ 
there is a crossover in the GS-structure. One particle becomes now totally 
free while the other two form a floppy pair with oscillatory motions. 
One can view this supersolid GS as a natural extension of the mean field 
HF state. Instead of having one particle in the mean field of the others, 
one has a pair in the field of the third particle. This is reminiscent 
of the BCS ansatz with a fixed number of fermions proposed in 
Ref. \cite{bouchaud}. Below $r_s \approx r_s^F$, the remaining 
pair melts and one recovers the HF mean field limit. 

 As explained in Sec. \ref{section2} for $N=2$, above the first threshold 
$r_s^F$, one has a continuous Wigner regime (oscillations of the 
molecule around the equilibrium positions exceeding the lattice spacing), 
before having important lattice effects at a second threshold 
$r_s^* \propto L^3$. As clear in Fig. \ref{FIG1} and Fig. \ref{fig61}, 
$L=6$ is too small to have the expected correlated three particle Wigner 
molecule in a continuous regime, without lattice effects. $L=6$ is just 
large enough for having a partially correlated supersolid regime free 
of significant lattice effects. However, the exact diagonalization study 
of $N=3$ spinless fermions can be extended to larger values of $L$. Such 
scaling analysis is in progress. The first results confirm the general 
picture emerging from this detailed study of the $L=6$ system, and support, 
at least for an ultimate mesoscopic limit, the possibility proposed by 
Andreev and Lifshitz for the thermodynamic limit: a quantum crystal may 
have delocalized defects without melting, the number of sites of the 
crystalline array being smaller than the total number of particles. 

On one side, one cannot exclude, as already pointed out in Ref. \cite{ksp}, 
that a supersolid regime is favored by the chosen geometry, because of 
the number of particles and underlying square lattice, but not favored at 
all in the continuous limit, which has not a square symmetry but a 
spontaneously broken hexagonal symmetry. On the other side, such a 
supersolid regime is not considered in the quantum Monte-Carlo (QMC) 
studies, since these methods rely on certain assumptions on the GS nodal 
structures. We study in App. \ref{Appendix C} those nodal structures 
in our small lattice model, which turn out to be very complex for 
intermediate values of $r_s$, and cannot be approximated by the nodal 
structures of the weak or the strong coupling limits. In 
App. \ref{Appendix D}, the GS occupation numbers in the reciprocal 
lattice are given when $r_s$ increases, and exhibit a hybrid structure 
in the mesoscopic supersolid regimes, the usual spreading of the ``Fermi 
sea'' being  accompanied by the gradual emergence of the Fourier 
spectrum of the ``Wigner solid''.

\begin{acknowledgement}
We thank Houman Falakshahi for stimulating discussions. 
Z. \'A. N\'emeth acknowledges the financial support provided 
through the European Community's Human Potential Programme under 
contract HPRN-CT-2000-00144 and the Hungarian Science Foundation 
OTKA TO34832.
\end{acknowledgement} 

\appendix
\section{Harmonic oscillatory motion of a two particle molecule}
\label{Appendix A}

In this appendix, the oscillatory motion of a two particle Wigner 
molecule on an empty $L \times L$ lattice with periodic BCs is studied 
when $r_s^F<r_s < r_s^*$. This corresponds to a Coulomb repulsion which 
is strong enough for forming a Wigner molecule, but not strong enough 
for restricting the oscillatory motion of the particles around equilibrium 
to scales of order of the lattice spacing $a$ (correlated lattice regime). 
As earlier noticed, the distance between two sites on a periodic lattice  
defined in Eq. \ref{distance} yields a cusp of the Coulomb repulsion at the 
equilibrium positions of the Wigner molecule. The role of this cusp 
is negligible in the Fermi limit, but not in the Wigner limit, the 
oscillatory motion of the particles around equilibrium being located 
at the cusp. To avoid this complication, we smear the long range part 
of the Coulomb repulsion, taking for the pairwise repulsion between two 
particles separated by $\bf{r}$ in the continuum limit of a system of size 
$D$
\begin{equation}
V_c({\bf r})=\frac{e^2 \pi}{D\sqrt{\frac{\sin^2\pi r_x}{D}+
\frac{\sin^2\pi r_y}{D}}}
\label{smeared}
\end{equation}
instead of the previous repulsion $V({\bf r})=e^2/|{\bf r}|$ with 
$|{\bf r}|$ defined by Eq. \ref{distance} for a square with periodic BCs. 
The corresponding two particle Hamiltonian
\begin{equation}
H_c=-{\hbar^2\over 2 m} (\nabla^2_1 + \nabla^2_2) + V_c({\bf r})
\label{contHam}
\end{equation}
is the sum of two decoupled terms
\begin{equation}
H_c=H_c({\bf R})+H_c({\bf r})
\end{equation}
when one uses the center of mass ${\bf R} = ({\bf r}_1 +{\bf r}_2)/2$ 
and relative separation ${\bf r}={\bf r}_1-{\bf r}_2$ coordinates. 
The Hamiltonian for the center of mass motion is given by 
\begin{equation}
H_c({\bf R}) = -{\hbar^2 \over 4 m} \nabla_{\bf R}^2
\end{equation}
while the Hamiltonian for the relative motion reads
\begin{equation}
H_c({\bf r})=-{\hbar^2 \over m} \nabla_{\bf r}^2 + U(\bf{r}).
\label{Harmham}
\end{equation}

\begin{figure}
{\centerline{\leavevmode \epsfxsize=4cm \epsffile{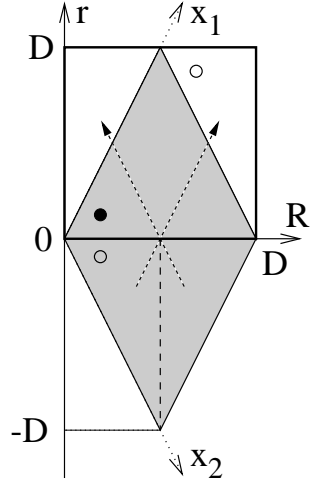}}}	
\caption{Re-definition of the domain of variation for $\bf R$ and $\bf r$ 
in one dimension. The gray area is the original domain 
$0<{\bf x}_1, {\bf x}_2<D$ with periodic BCs. Moving the part where 
$R<0$ to the upper triangles, the new domain becomes $0< \vec{R,r} <D$ 
with periodic BCs. The bottom left empty circle, which is the 
$x_1 \leftrightarrow x_2$ exchange counterpart of the solid circle, 
goes to the empty circle in the upper right triangle.}
\label{figa1}
\end{figure}

It is convenient to redefine the domain where $\bf R$ and $\bf r$ vary 
such that it becomes again a $2d$ torus. Fig. \ref{figa1} 
shows how it can be done in one dimension. This change of domain modifies 
the usual symmetry requirement 
\begin{equation}
\Psi(R,r)=\Psi(x_1,x_2)=-\Psi(x_2,x_1)=-\Psi(R,-r), 
\end{equation} 
for spinless fermions, since $\psi(R,-r)\rightarrow \Psi(R+D/2,-r+D)$.
In two dimensions, the symmetry requirement for spinless fermions after 
this change of domain becomes:
\begin{equation}
\Psi({\bf R},{\bf r})=-\Psi\left({\bf R}+\left({D\over 2},{D\over 2}\right),-{\bf r}+(D,D)\right). 
\label{sf-symmetry}
\end{equation}
The eigenstates of $H_c=H_c({\bf R})+H_c({\bf r})$ take the form:  
\begin{equation}
\Psi({\bf R},{\bf r}) = {1\over D}e^{i {\bf KR}} \psi({\bf r}),  
\end{equation}
the symmetry requirement for $\psi({\bf r})$ being:
\begin{equation}
\psi({\bf r})=\cases{-\psi\Bigl((D,D)-{\bf r}\Bigr),&
for $\frac{D}{2 \pi} (K_x+K_y)$ even;
\cr\psi\Bigl((D,D)-{\bf r}\Bigr),& for $\frac{D}{2 \pi} (K_x+K_y)$ odd.} 
\label{fermion}
\end{equation}

When a sufficient Coulomb repulsion yields small oscillation of the 
inter-particle spacing around its largest possible value, the Coulomb 
repulsion can be expanded and one gets for the relative motion of the 
two particles a $2d$ harmonic oscillator Hamiltonian:
\begin{equation}
H_c({\bf r})=-{\hbar^2 \over m} \nabla_{\bf r}^2 + {e^2 \pi\over D \sqrt{2}}+
{e^2\pi^3 \over D^3 \sqrt{32}} \left|{\bf r}-\left({D\over 2},{D\over 2}\right)\right|^2.
\end{equation}

 For a $2d$ harmonic oscillator 
\begin{equation}
H_{osc}=-{\hbar^2\over 2m} \nabla_{\bf r}^2 + {m \omega ^2 \over 2} 
|\vec{r}|^2, 
\end{equation}
the symmetric and antisymmetric GSs of energies 
$E_{0S}=\hbar \omega$ and $E_{0A}=2 \hbar \omega$ are 
given by:
\begin{equation}
\Psi_{0S}({\bf r})=\frac{1}{l_0\sqrt{\pi}} 
\exp \left(- \frac{|{\bf r}|^2}{2l_0^2} \right)
\end{equation}
and
\begin{equation}
\Psi_{0A}(\vec{r})=\frac{r_x\sqrt2}{l_0^{3/2}\sqrt{\pi}} 
\exp \left(- \frac{|\vec{r}|^2}{2l_0^2} \right)
\end{equation}
respectively. 
The length $ l_0=\sqrt{{\hbar}/({m\omega})}$
becomes in our case 
\begin{equation}
l_0 = \left( {\hbar^2 \over m} {\sqrt{32} D^3 \over e^2 \pi^3} \right)^{-1/4}.
\end{equation}

Discretizing the continuous Hamiltonian on $L\times L$ lattice, one gets  
\begin{eqnarray}
H_l&=&-t \sum_{\left<{\bf j},{\bf j'}\right>}c_{\bf j}^\dagger c_{\bf j'}
+4Nt \nonumber\\ 
&& + {U \pi\over L\sqrt{\sin^2{(j_{x1}-j_{x2})\pi\over
L}+\sin^2{(j_{y1}-j_{y2})\pi\over D}}}, 
\end{eqnarray}
where $L= D / a$,  $U=e^2/a$ and $t=\hbar^2 /2 m a^2$. The constant term 
$4Nt$ comes from the discretization of the 
Laplacians. For $H_l$, the characteristic length $l_0$ becomes  
\begin{equation}
l_0 = \left( \frac{8 \sqrt{2}t}{U L \pi^3} \right)^{-1/4} D. 
\end{equation}

\begin{figure}[t]
{\centerline{\leavevmode \epsfxsize=9cm \epsffile{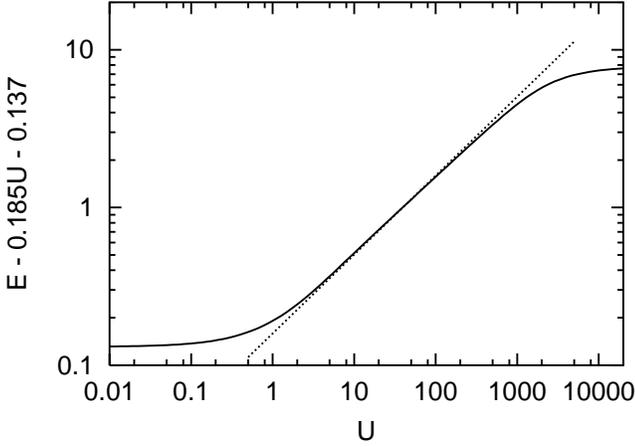}}}	
\caption
{GS energy $E_0$ for two particles in a $12 \times 12$ lattice as 
a function of $U$. We have plotted for convenience $E=E_0-0.185 U -0.137$ 
(taking $t=1$) for checking the contribution $\sim U^{1/2}$ at intermediate $U$ 
of Eq. \ref{eos}. The analytical estimate ($0.159\sqrt{U}$, dotted line) 
is valid for $1<U<1000$, defining the continuous Wigner regime where 
the relative fluctuation $u_r\propto r_s^{-1/4}$ (Fig \ref{FIG1}).
}
\label{fig22}
\end{figure}

Since $l_0 \propto r_s^{-1/4} D $ in our model, one gets 
$\left<r\right> ={\sqrt 2} D$ for the average inter-particle spacing 
and $\Delta r \propto r_s^{-1/4} D $ for the width of its 
distribution. Therefore, the ratio $u_r$ of these two quantities decays 
as $r_s^{-1/4}$. This corresponds to the behaviors shown in 
Fig. \ref{FIG1} when $r_s^F < r_s <r_s^*$.

For a GS of total momentum ${\bf K}=2\pi/D(1,0)$, the condition 
(\ref{fermion}) yields the symmetric GS for $ H_c({\bf r})$. 
For the two particle GS of  $ H_c({\bf R}) + H_c({\bf r})$, 
one gets an energy 
\begin{equation}
E_0={\hbar^2\over 4m} {\bf K}^2+{e^2\pi\over D\sqrt{2}}+\hbar \omega,
\end{equation}
where
\begin{equation}
\hbar \omega = \sqrt{{4 \hbar^2\over m} {1\over2}{e^2 \pi^3\over D^3
\sqrt{8}}}.
\end{equation}
This energy becomes for the corresponding lattice Hamiltonian $H_l$:
\begin{equation}
E_{0}={2\pi^2\over L^2}t+{U \pi\over L \sqrt{2}}+\sqrt{{Ut\over L^3}{4
\pi^3\over\sqrt{8}}}. 
\label{eos}
\end{equation}
 A numerical check of this expression using a $12 \times 12$ lattice model 
is shown in Fig. \ref{fig22}. Dividing $E_{0}$ by $U^2/t$, one gets 
\begin{equation}
\frac{E_0 t}{U^2}= \frac {\pi}{4 r_s^2} + \frac{\sqrt{\pi}}{4 r_s} + 
\frac{\pi^{3/4}}{4 r_s^{3/2}}.
\end{equation}
This expansion is similar to the original expression given by Wigner 
\cite{wigner} for the strong coupling limit, the energy being measured in 
Rydberg units. The first term gives the kinetic energy of the center of 
mass ($\sim r_s^{-2}$), the second is the electrostatic energy at 
equilibrium ($\sim r_s^{-1}$) and the third term comes from the oscillations 
of the inter-particle spacing around equilibrium ($\sim r_s^{-3/2}$). 

\section{Correlated lattice limit when $L \rightarrow \infty$} 
\label{Appendix B}

\begin{figure}[t]
{\centerline{\leavevmode \epsfxsize=6cm \epsffile{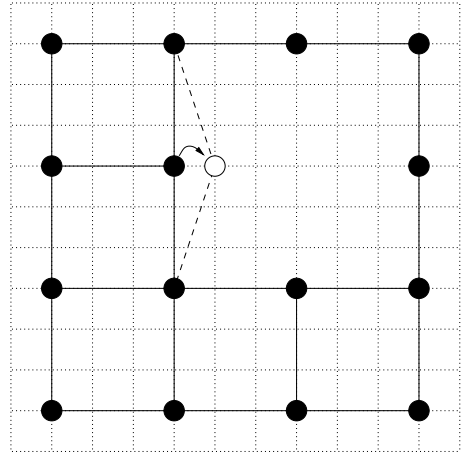}}}	
\caption
{
$N=n^2-1$ spinless fermions on a square lattice 
of size $L=3n$ for $n=4$. One of the $L^2$ square Wigner lattices 
with a vacancy of energy $E_0^{(0)}$ (solid circles) and one virtual 
state of energy $E_1^{(0)}$ (empty circle) contributing to the vacancy 
dynamics are indicated.
}
\label{figapp}
\end{figure}

We study two limits which can be easily described by the $t/U$ lattice 
perturbation theory when $t \rightarrow 0$ and $L \rightarrow \infty$. 

We keep $N=3$ spinless fermions in the first limit. In this case, the 
hopping term $\sim h$ characterizing the rigid translation of the 
molecule remains of order $t^3/U^2$, while the hopping term $\sim r$ 
characterizing a single particle hop over $L/2$ sites and coupling 
triangles of different orientations becomes of order $t (t/U)^{L/2-1}$. 
When $L \rightarrow \infty$, only the rigid translation of the triangle 
matters, and the effective Hamiltonian (Eq. \ref{effham3}) reads:
\begin{equation}
H_{eff}^{(III)}=\sum_{\bf j} E_0^{(II)} D_{\bf j}^\dagger D_{\bf j}+h
\sum_{\left<{\bf j},{\bf j'}\right>} D_{\bf j}^\dagger D_{\bf j'},
\end{equation}
where $D^\dagger_{\bf j}$ ($D_{\bf j}$) creates (annihilates) a triangle 
defined by Eq. \ref{cre3}, replacing $3$ by $L/2$. The resulting energies 
are:
\begin{equation}
E^{(III)}_{\bf K} = {E}_0^{(II)} + 2h\left(\cos K_x+\cos K_y\right), 
\end{equation}
yielding a GS total momentum ${\bf K}=(0,0)$.

\begin{figure}[b]
{\centerline{\leavevmode \epsfxsize=9cm \epsffile{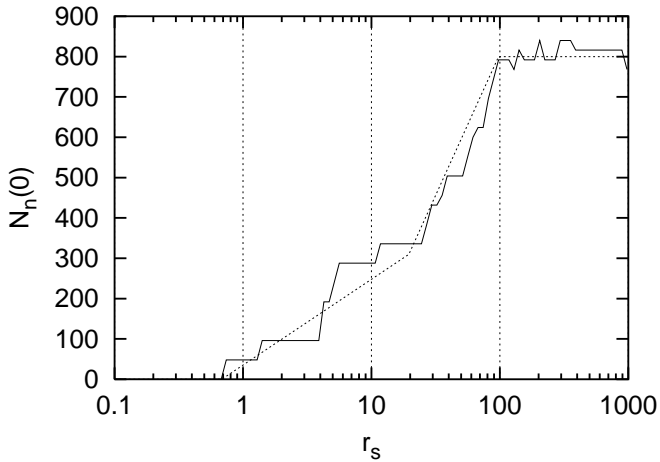}}}	
\caption{
The number of negative components $N_n(0)$ as a function 
of $r_s$. The solid line is obtained with a threshold $\zeta =10^{-12}$. 
The dotted line is a guide to the eyes.} 
\label{fig81}
\end{figure}

We take $N=n^2-1$ spinless fermions and a size $L=3n$ in the second limit. 
$N=n^2$ gives a uniform filling factor $\nu=1/9$ and a square Wigner lattice 
which is commensurate with the assumed lattice. Taking a single particle out 
$n^2$ will create a single vacancy in the square Wigner lattice, as shown 
in Fig. \ref{figapp}. In this second limit, this is now the rigid 
translation of the Wigner lattice which becomes negligible in the 
thermodynamic limit, while the hopping term $r$ characterizing the 
propagation of the vacancy remains $\propto t^3/U^2$. The effective 
Hamiltonian (Eq. \ref{effham3}) takes the form:
\begin{equation}
H_{eff}^{(III)}=\sum_{\bf j} E_0^{(II)} D_{\bf j}^\dagger D_{\bf j}+
r \sum_{\left<{\bf j},{\bf j'}\right>_3}D_{\bf j}^\dagger D_{\bf j'}.
\end{equation}
There is only a set of virtual states of energy $E_1^{(0)}$ which contribute 
at the lowest order to the propagation of the vacancy (see 
Fig. \ref{figapp}), making simple to calculate $r$.  After Fourier 
transformation, the eigenenergies of the effective Hamiltonian are given 
by:
\begin{equation}
E^{(III)}_{\bf K}=E_0^{(II)}+ {2t^3\left(\cos 3 K_x + \cos 3 K_y 
\right)\over \left(E_1^{(0)}-E_0^{(0)} \right)^2},
\end{equation}
and corresponds to the spectrum of a single particle on the 
assumed square lattice with third nearest neighbor hopping. Of course, 
this one vacancy dynamics does not totally remove the $L^2$ GS degeneracy 
of the limit $t \rightarrow 0$, but makes very simple the study of charge 
propagation in this highly correlated many particle system. 

\section{Nodal structure of the three particle system}
\label{Appendix C}

\begin{figure}[b]
{\centerline{\leavevmode \epsfxsize=9cm \epsffile{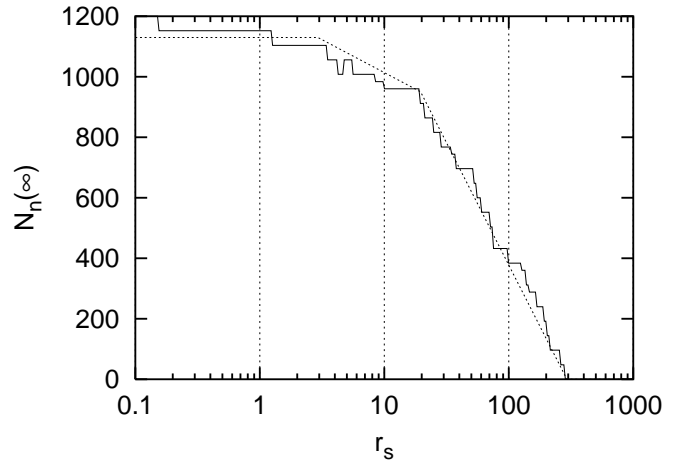}}}	
\caption{
The number of negative components $N_n(\infty)$ as a function 
of $r_s$. The solid line is obtained with a threshold $\zeta =10^{-15}$. 
The dotted line is a guide to the eyes.}
\label{fig82}
\end{figure}

\begin{figure}[t]
{\centerline{\leavevmode \epsfxsize=9cm \epsffile{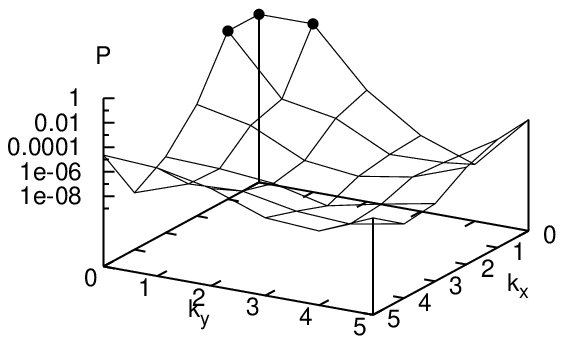}}}
{\centerline{\leavevmode \epsfxsize=9cm \epsffile{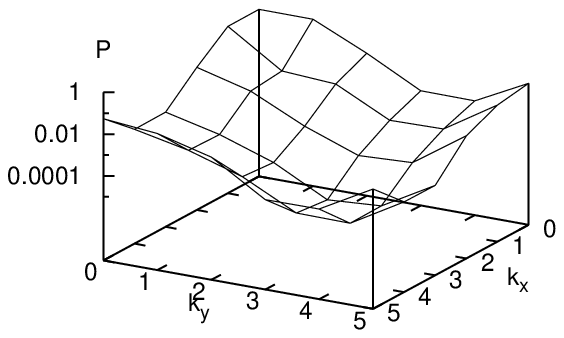}}}	
{\centerline{\leavevmode \epsfxsize=9cm \epsffile{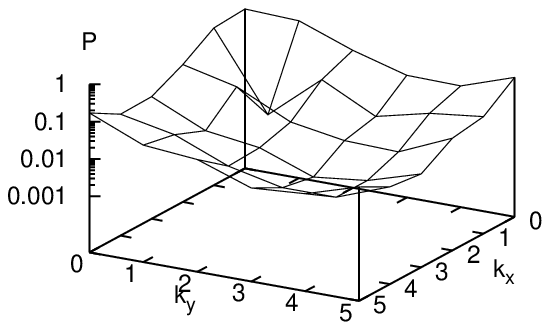}}}	
\caption{Occupation numbers $P_{\bf k}$ on the reciprocal lattice 
($2 \pi k_x/L,2 \pi k_y/L $) for $r_s=0.1$ (upper figure), $r_s=10$ 
(middle figure) and $r_s=40$ (lower figure).}
\label{fig71}
\end{figure}

   The quantum Monte-Carlo (QMC) methods are the most powerful tools 
for studying large many-body systems \cite{ceperley}. However, the 
study of the ground state of fermionic systems suffers from the 
well known ``sign problem''. One way to avoid the negative weights 
that would be otherwise generated by antisymmetric states is the 
fixed node approximation \cite{ceperley}. The fixed node GS energy 
is then an upper bound to the exact GS energy. The nodal structure of 
the liquid limit is given by a simple Slater determinant of plane waves, 
and of localized orbitals for the solid limit. To know the exact nodal 
structure for intermediate couplings is not an easy problem 
\cite{ceperley-node}. In this appendix, we study the nodal structure 
of three spinless fermions on a $6 \times 6$ lattice. 

 Previously, we have considered ${\bf K}$-eigenstates, for an 
interacting system which is invariant under lattice translations. The 
Hamiltonian being invariant under time-reversal symmetry, one first 
define eigenvectors with real components in the site basis. For this 
purpose, we combine the $\bf K$-eigenvector
\begin{equation}
\left|\Psi_{\bf K}(r_s)\right>=\sum_{{\bf k}_1,{\bf k}_2,{\bf k}_3 
\atop {\bf k}_1+{\bf k}_2+{\bf k}_3={\bf K}} 
\Psi_{{\bf k}_1,{\bf k}_2,{\bf k}_3}(r_s)
c^\dagger_{{\bf k}_1} c^\dagger_{{\bf k}_2} c^\dagger_{{\bf k}_3} 
\left|0\right>,
\end{equation}
with its time reversed conjugate $-\bf K$-eigenvector:
\begin{equation}
\left|\Psi_{-{\bf K}}(r_s)\right>=\sum_{{\bf k}_1,{\bf k}_2,{\bf k}_3 
\atop {\bf k}_1+{\bf k}_2+{\bf k}_3={\bf K}} 
\Psi_{{\bf k}_1,{\bf k}_2,{\bf k}_3}(r_s)
c^\dagger_{-{\bf k}_1} c^\dagger_{-{\bf k}_2} c^\dagger_{-{\bf k}_3} 
\left|0\right>.
\end{equation}
Since the $\Psi_{{\bf k}_1,{\bf k}_2,{\bf k}_3}(r_s)$ are real, the 
combination  
\begin{equation}
\Psi_{{\bf j}_1,{\bf j}_2,{\bf j}_3}(r_s)= {1\over\sqrt{2}} \left<0\right| 
c_{{\bf j}_3} c_{{\bf j}_2} c_{{\bf j}_1}
\Bigl(\left|\Psi_{\bf K}(r_s)\right>+\left|\Psi_{-{\bf K}}(r_s)\right>\Bigr)
\end{equation}
is indeed real in the site basis. 

We want to compare the nodal structure of this real GS at intermediate 
values of $r_s$ to the two limiting nodal structures, characterizing the 
GS either in the limit $r_s \rightarrow 0$ or $r_s \rightarrow \infty$. We 
define the components of a vector $\chi (r_s,lim)$ by 
\begin{equation}
\chi_{{\bf j}_1,{\bf j}_2,{\bf j}_3}(r_s,lim)
=\Psi_{{\bf j}_1,{\bf j}_2,{\bf j}_3}(r_s) 
\cdot \Psi_{{\bf j}_1,{\bf j}_2,{\bf j}_3}(lim);
\end{equation}
where $\Psi(lim)$ is the corresponding limiting GS, and we count the 
number $N_n (lim)$ of negative components of $\chi$. 
If the nodal structures of the limit and at $r_s$ are identical, 
$N_n(lim)=0$. However, when $\chi_{{\bf j}_1,{\bf j}_2,{\bf j}_3}(r_s,lim)$ 
is almost zero, its sign becomes undefined due to numerical precision. 
This is why we ignore all the components of $\chi$ below a given threshold 
$\zeta$. The behavior of $N_n (0)$ ($N_n(\infty)$) where $\Psi(lim)$ is a 
liquid GS $\Psi(r_s=0.1)$ (solid GS $\Psi(r_s=10^3)$) is shown in 
Fig. \ref{fig81} (Fig. \ref{fig82}).

As one can see, the nodal structure does not exhibit a sharp transition 
between the two limits, but a crossover with complex intermediate behaviors. 
Notably, there are some plateau values suggesting some constant nodal 
structure around the intermediate values of $r_s$ where we observe the 
PMWM states. This illustrates the difficulty to use a fixed node 
Monte-Carlo method for describing the intermediate GS on a lattice. 

\begin{figure}[t]
{\centerline{\leavevmode \epsfxsize=9cm \epsffile{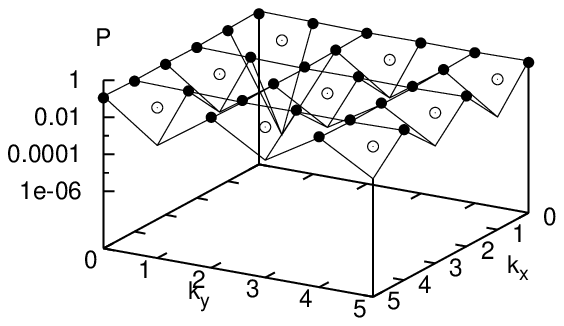}}}	
{\centerline{\leavevmode \epsfxsize=5cm \epsffile{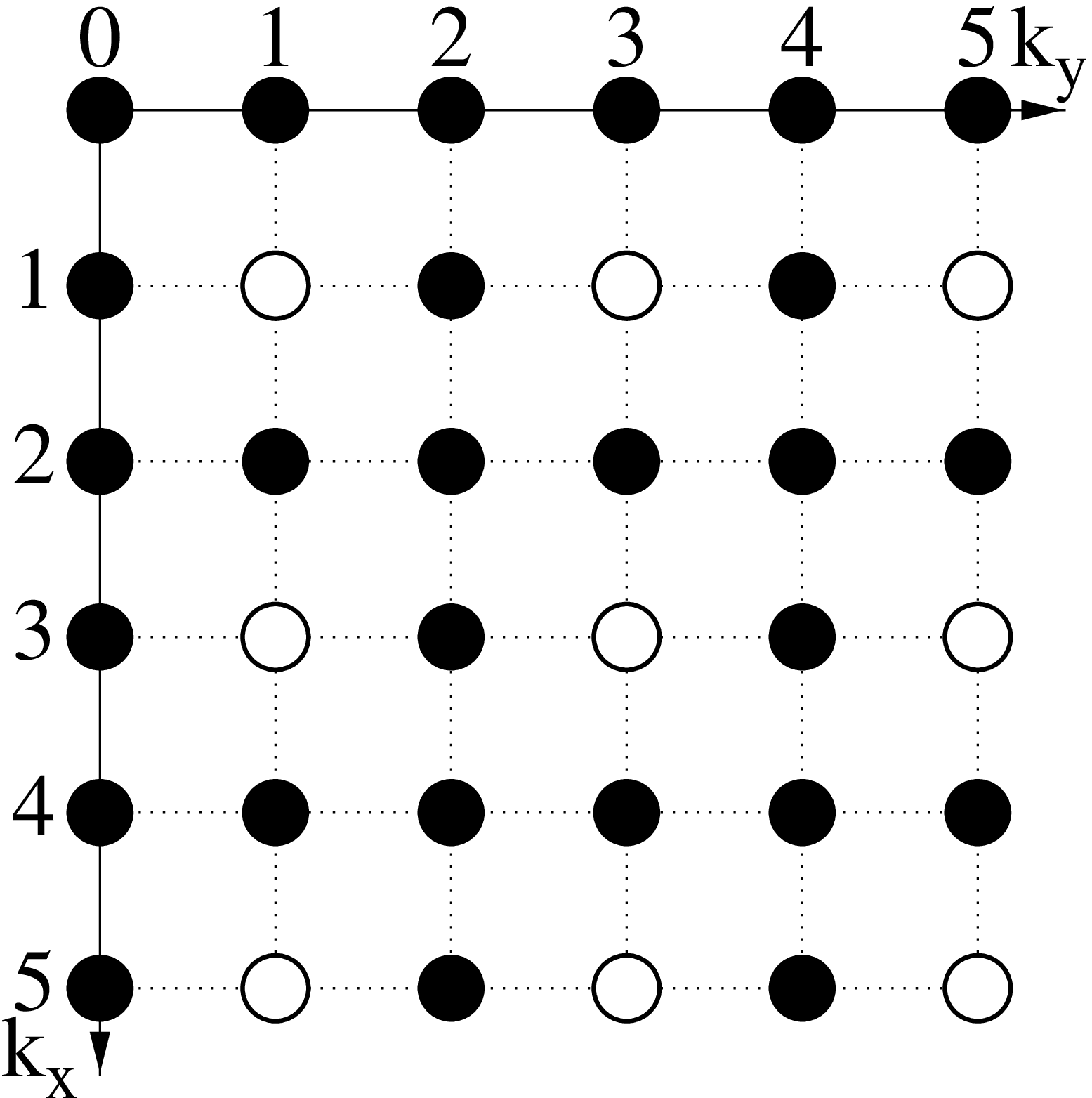}}}	
\caption{{\bf Above:} Occupation numbers $P_{\bf k}$ for $r_s=500$. 
{\bf Below:} Occupation numbers $P_{\bf k}$  at $r_s=\infty$, showing 
the two possible values $1/9$ (filled circles) and $0$ (empty circles).}
\label{fig74}
\end{figure}

\section{Occupation numbers in ${\bf k}$-space}
\label{Appendix D}

 It is not only interesting to know how the system occupies the real 
lattice, but also the reciprocal lattice. From the GS wave function 
$\Psi_0(r_s)$ of total momentum  ${\bf K}=2 \pi/6 (1,1)$, we have 
calculated the occupation numbers
\begin{equation}
P_{\bf k}(r_s)=\left< \Psi_0(r_s) \right| d^\dagger_{\bf k} d_{\bf k} 
\left|\Psi_0(r_s)\right>,
\end{equation}
in ${\bf k}$-space. Fig. \ref{fig71} and Fig. \ref{fig74} give 
plots of those number in the reciprocal lattice for increasing 
values of $r_s$. When $r_s \rightarrow 0$, only three ${\bf k}$-states 
are occupied (Fig. \ref{fig71}). When $r_s \rightarrow \infty$, 
one gets a simple pattern (Fig. \ref{fig74}) for ${\bf K}=2\pi/6(1,1)$, 
which can be analytically obtained:
\begin{equation}
P_{\bf k} (r_s=\infty)=\cases{0 &if ${\bf k}=2\pi/6 \times$(odd,odd) 
\cr 1/9 & else}.
\end{equation}

 Between the two limits, the occupation numbers $P_{\bf k}$ of the 
two identified PMWMs ($r_s=10$ and $r_s=40$) have a mixed character, 
where both the ``solid'' and the ``liquid'' patterns are visible 
when one uses a logarithmic scale. This is what one should expect 
when a ``supersolid'' is forming: the usual spreading of the ``Fermi 
sea'' being  accompanied by the gradual emergence of the Fourier 
spectrum of the ``Wigner solid''.

\end{document}